\documentclass[a4paper,12pt]{article}
\usepackage{graphicx}
\usepackage{psfrag}

\usepackage{mathrsfs}
\usepackage{amssymb}
\usepackage{amsfonts}
\usepackage{latexsym}
\usepackage{amsmath}
\usepackage{amsthm}
\usepackage[cp1251]{inputenc}
\usepackage[english]{babel}

\newcommand{\er}[1]{\textrm{(\ref{#1})}}
\def\lb{\label}

\theoremstyle{plain}

\newtheorem{theorem}{\bf Theorem}[section]
\newtheorem{lemma}[theorem]{\bf Lemma}

\theoremstyle{remark}

\renewcommand{\a}{\alpha}

\newcommand{\g}{\gamma}           

\newcommand{\G}{\Gamma}           \newcommand{\cD}{\mathcal{D}}

\renewcommand{\d}{\delta}           

\newcommand{\D}{\Delta}           

\newcommand{\ve}{\varepsilon}

\newcommand{\e}{\eta}             

\newcommand{\vt}{\vartheta}       

\newcommand{\vT}{\Theta}

\renewcommand{\l}{\lambda}          \newcommand{\cM}{\mathcal{M}}

\renewcommand{\L}{\Lambda}

\newcommand{\n}{\nu}              

\renewcommand{\r}{\rho}             

\newcommand{\s}{\sigma}           \newcommand{\cR}{\mathcal{R}}

\renewcommand{\t}{\tau}             

\newcommand{\f}{\phi}             

\newcommand{\F}{\Phi}             

\newcommand{\vp}{\varphi}

\newcommand{\p}{\psi}             
 
\renewcommand{\P}{\Psi}             
      
\renewcommand{\o}{\omega}
\renewcommand{\O}{\Omega}
\newcommand{\x}{\xi}

\newcommand{\vk}{\varkappa}

\newcommand{\gS}{\mathfrak{S}}

\def\Z{\mathbb{Z}}
\def\R{\mathbb{R}}
\def\C{\mathbb{C}}

\def\N{\mathbb{N}}

\def\qqq{\qquad}
\def\qq{\quad}

\let\ge\geqslant
\let\le\leqslant

\newcommand{\ca}{\begin{cases}}
\newcommand{\ac}{\end{cases}}
\newcommand{\ma}{\begin{pmatrix}}
\newcommand{\am}{\end{pmatrix}}

\def\lt{\biggl}
\def\rt{\biggr}

\renewcommand{\[}{\begin{equation}}
\renewcommand{\]}{\end{equation}}

\def\wt{\widetilde}
\def\pa{\partial}
\def\sm{\setminus}
\def\es{\emptyset}
\def\no{\noindent}
\def\ol{\overline}
\def\iy{\infty}

\def\/{\over}
\def\ts{\times}

\def\ss{\subset}

\def\Re{\mathop{\rm Re}\nolimits}

\def\Im{\mathop{\rm Im}\nolimits}

\def\Tr{\mathop{\rm Tr}\nolimits}

\def\BBox{\hspace{1mm}\vrule height6pt width5.5pt depth0pt \hspace{6pt}}
\def\wh{\widehat}

\begin{document}

\title {Spectral estimates for
periodic fourth order operators  }

\author{Andrey Badanin
\begin{footnote}
{ Department of
 Mathematics, Arkhangelsk State Technical Univ.,
 Severnaya Dvina emb., 17, 163002, Arkhangelsk, Russia,
 e-mail: a.badanin@agtu.ru}
\end{footnote}
 \and Evgeny Korotyaev
\begin{footnote}
{ School of Mathematics, Cardiff Univ., Senghennydd Road, Cardiff,
CF24 4AG, UK, e-mail: KorotyaevE@cf.ac.uk}
\end{footnote}
}

\maketitle

\begin{abstract}
\no We consider the operator $H={d^4\/dt^4}+{d\/dt}p{d\/dt}+q$ with
1-periodic coefficients on the real line.
The spectrum of $H$ is absolutely continuous and consists of intervals  separated by gaps.
We describe the spectrum of this operator in terms of
the Lyapunov function, which is analytic on a  two-sheeted Riemann
surface.
On each sheet the Lyapunov function has the standard
properties of the Lyapunov function for the scalar case.
 We describe the spectrum of $H$ in terms of periodic, antiperiodic eigenvalues,
and so-called resonances.
 We prove that 1) the spectrum of $H$ at high energy has
multiplicity two, 2)  the asymptotics of the periodic,
antiperiodic eigenvalues and of the resonances are  determined at high energy, 3) for some specific $p$ the spectrum of $H$ has an infinite number of gaps, 4)
the spectrum of $H$ has small spectral band (near the beginner of
the spectrum) with multiplicity $4$ and its asymptotics are
determined as $p\to 0, q=0$.

\end{abstract}

\section {Introduction and main results}
\setcounter{equation}{0}

We consider the self-adjoint operator $H={d^4\/dt^4}+{d\/dt} p
{d\/dt}+q,$ acting in $L^2(\R)$, where the real  coefficients $p,q$
are 1-periodic and $p,p',q \in L^1(0,1)$. Here and below we use the notation $f'={d f\/d t},
f^{(k)}={d^kf \/d t^k}$. The spectrum $\s(H)$ of $H$ is absolutely continuous and consists of
non-degenerated intervals (see \cite{DS}). These intervals are separated by gaps
with length $\ge 0$. Introduce the fundamental solutions
$\vp_j(t,\l), j\in\N_3^0=\{0,1,2,3\}$  of the equation
\[
\lb{1b} y''''+( p y')'+q y=\l y,\qqq (t,\l) \in \R\ts\C,
\]
satisfying the conditions: $\vp_j^{(k)}(0,\l)=\d_{jk},j,k=\N_3^0$,
where $\d_{jk}$ is the standard Kronecker symbol. We define the
monodromy $4\ts 4$-matrix $M$ and its characteristic polynomial $D$
by
\[
\lb{1c} M(\l)=\{\vp_j^{(k)}(1,\l)\}\}_{k,j=0}^3,\qq
D(\t,\l)=\det(M(\l)-\t I_4),\qq \t,\l\in\C.
\]
The matrix valued function $M$ is entire and real on $\R$.
An eigenvalue $\t(\l)$ of
$M(\l)$ is called a {\it multiplier}, i.e., it is a zero of the
algebraic equation $D(\t,\l)=0$. It is known that if $\t(\l)$  is a
multiplier of multiplicity $d$ for some $\l\in\C$ (or $\l\in\R$),
then $\t^{-1}(\l)$ (or $\ol\t(\l)$) is a multiplier of multiplicity
$d$ (see \cite{T1}). Moreover, each $M(\l), \l\in\C$ has exactly
four multipliers $\t_1^{\pm1}(\l), \t_2^{\pm1}(\l)$. Furthermore,
$\s(H)=\{\l\in \C:|\t_1(\l)|=1 \ or \ |\t_2(\l)|=1\}$. Now we
formulate our preliminary result about eigenvalues of the monodromy
matrix. Below we use $z=\l^{1\/4},\arg z\in(-{\pi\/4},{\pi\/4}]$ for
$\l\in\C$.

\begin{theorem}
\lb{T1}
Let $\t_\n ,\t_\n^{-1},\n=1,2$ be eigenvalues of $M$. Then

\no i) The Lyapunov functions $\D_\n={1\/2}(\t_\n+\t_\n^{-1}),\n=1,2$,
are branches of $\D=T_1+\sqrt{\r}$ on the two sheeted Riemann
surface $\cR$ defined by $\sqrt\r$ and satisfy
\[
\lb{1d}
\D_\n=T_1-(-1)^\n\sqrt{\r},\qq
\r={T_2+1\/2}-T_1^2,\qq
T_\n={1\/4}\Tr M^\n,
\]
\[
\lb{2l} D(\t,\cdot)
=\det(M-\t I_4)
=(\t^2-2\D_1\t+1)
(\t^2-2\D_2\t+1),\qq \t\in\C,
\]
\[
\lb{aD1} \D_1(\l)=\cosh z(1+O(1/z))\qq\text{as}\qq |\l|\to\iy,\qq
|z-(1\pm i)\pi n|>1,\qq n\ge 0,
\]
\[
\lb{aD2} \D_2(\l)=\cos z(1+O(1/z))\qq\text{as}\ \ |\l|\to\iy,\ \
|z-(1\pm i)\pi n|>1,\ \ |z-\pi n|>1,\ \ n\ge 0.
\]

\no ii) If $\D_\n(\l)\in(-1,1)$ for some $(\n,\l)\in\{1,2\}\ts\R$
and $\l$ is not a branch point of $\D$,
then $\D_\n'(\l)\ne 0$.

\no iii) The following identity holds:
\[
\lb{sp}
\s(H)=\s_{ac}(H) =\{\l\in\R:\D_\n(\l)\in[-1,1]\ \text{for\ some}\
\n=1,2\}.
\]
Let $s_\n=\{\l\in\R:\D_\n(\l)\in[-1,1]\},\n=1,2$. Then the spectrum
of $H$ in the set $s_1\cap s_2$ has multiplicity $4$, and the
spectrum in the set $\s(H)\sm(s_1\cap s_2)$ has multiplicity $2$.
\end{theorem}

\no {\bf Remark.} 1) The functions $T_1,T_2,
\r,D_\pm={1\/4}\det(M\mp I_4)$ are entire and real on $\R$.

\no 2) Asymptotics \er{ero} show that $\r>0$ on  $(r,+\iy)$ for some $r\in\R$.
For the $2\ts 2$ matrix Schr\"odinger operator
the corresponding function may be equal to $0$ (see \cite{BBK}).

\no 3)  Theorem \ref{T1} is standard for the Schr\"odinger operator  with
$2\ts2$ matrix-valued potential \cite{BBK}, for the Dirac system
with $4\ts4$ matrix-valued potential \cite{K}.
Some similar results for the periodic Euler-Bernoulli
equation $(ay'')''=\l b y$ see in \cite{P1},\cite{P2},\cite{PK},
and for the operator $H$  see in \cite{T1}, \cite{T2}.

The zeros of $D_+($or $ D_-)$ are periodic (or antiperiodic)
eigenvalues for the equation \er{1b}. Due to \er{2l}, they are zeros
of $\D_\n -1$ (or $\D_\n +1$) for some $\n =1,2$.
Denote by $\l_0^+,\l_{2n}^\pm, n=1,2,...$
the sequence of zeros of $D_+$ (counted with multiplicity) such that
$\l_{0}^+\le \l_{2}^-\le \l_{2}^+\le \l_{4}^-\le\l_{4}^+\le \l_{6}^-
\le... $ Denote by $\l_{2n-1}^\pm, n=1,2,...$ the sequence of zeros
of $D_-$ (counted with multiplicity) such that $\l_{1}^-\le
\l_{1}^+\le \l_{3}^-\le \l_{3}^+\le\l_{5}^-\le \l_{5}^+ \le... $.

We call the zero of $\r$ the resonance.
The resonances of odd multiplicity are branch points
of the Lyapunov function $\D$. The resonances can be real
and non-real (see \cite{BK}).
The function $\r$ is real on $\R$, then
$r$ is a zero of $\r$ iff $\ol r$ is a zero of $\r$.
By Lemma \ref{ar} iii),
$\r$ has an odd number of real zeros
(counted with multiplicity) on the interval
$(-\G,\G)\ss\R,\G=4(\pi(N+{1\/2}))^4$ for sufficiently large $N\ge 1$.
Let $r_0^-,r_{n}^\pm, n\in \N$ be the sequence of
all zeros of $\r$ in $\C$ (counted with multiplicity) such that:

\no $r_0^-$ is a maximal real zero and $r_n^+\in\ol{\C_-}$,
$...\le\Re r_{n+1}^+\le\Re r_{n}^+\le...\le\Re r_1^+$,

\no if $r_n^+\in\C_-$, then $r_n^-=\ol{r_n^+}\in\C_+$,

\no if $r_n^+\in\R$, then $r_n^-\in\R$ and
$r_n^-\le r_n^+\le\Re r_{m-1}^-,m=1,...,n$.

\no Let $...\le r_{n_j}^-\le r_{n_{j-1}}^+\le...\le r_{n_3}^-\le
r_{n_2}^+ \le r_{n_1}^-\le r_{n_1}^+ \le r_{0}^-,j\ge 1,$ be the
subsequence of all real zeros of $\r$. Then  $\r(\l)<0$  for any
$\l\in(r_{n_{j+1}}^+,r_{n_j}^-),j\ge 1$.

Note that if $p=q=0$, then the corresponding functions have
the forms
\[
\lb{0} T_\n^0={\cosh \n z +\cos \n  z\/2},\qq \r^0={(\cosh z-\cos
z)^2\/4},\qq D_\pm^0=(\cos z\mp 1)(\cosh z\mp 1),
\]
and $\D_1^0=\cosh z, \D_2^0=\cos z$. Moreover, the identities \er{0} give $\r^0\le 0$ on $\R_-$
 and the corresponding resonances $r_0^-=0$, $r_n^\pm=-4(\pi n)^4$.


We describe the spectrum in terms of the Lyapunov function.

\begin{theorem}\lb{bands}
 i) For each $n\ge 1$ there exist 3 cases:

$i_1$) the function  $\D$ is real analytic and $\D'\ne 0$ on
$\s_n'=(\l_{n-1}^+, \l_{n}^-)$, and $\D(\s_n')\ss(-1,1)$,

$i_2$) there exists a zero $\wt\l_n^-$
of $\r$ such that
$\wt\l_n^-<\min \{\l_{n-1}^+, \l_{n}^-\}$ and one branch of $\D$
 is real analytic and its derivative
$\ne 0$ on $\s_n^-=(\wt\l_n^-,\l_{n-1}^+)$
and another branch of $\D$
 is real analytic and its derivative $\ne 0$ on $\s_n^+=(\wt\l_n^-,\l_{n}^-)$,

$i_3$) there exists a zero $\wt \l_n^+$ of $\r$
such that
$\wt\l_n^+>\max \{\l_{n-1}^+, \l_{n}^-\}$ and one branch of $\D$
is real analytic and its derivative $\ne 0$
on $\s_n^-=(\l_{n-1}^+,\wt\l_n^+)$,
and another branch of $\D$
is real analytic and its derivative $\ne 0$ on
$\s_n^+=(\l_{n}^-,\wt\l_n^+)$.

Moreover,  in the cases $i_2),i_3)$   let $\s_n'=\s_n^-\cup\s_n^+$, then
$\D(\s_n')\ss (-1,1)$, furthermore, the zero $\wt\l_n^{\pm}$ of $\r$ has
multiplicity $1$ (it is a branch point of $\D$) or $2$.

\no ii) The spectrum $\s(H)$ satisfies:
\[
\lb{spH}
\s(H)=\cup_{n\ge 1}\s_n,\qq \s_n=\ol{\s_n'},\qqq
\s_n\cap\s_{n+2}=\es,\qq n\ge 1.
\]
The spectrum $\s(H)$ of $H$ has multiplicity $4$ in the set
\[
\lb{s4}
\gS_4=\lt(\bigcup_{n\ge 1}(\s_n'\cap\s_{n+1}')\rt)\bigcup
\lt(\bigcup_{n:\ i_2)\ \text{or}\ i_3)\ \text{holds}}
(\s_n^-\cap\s_n^+)\rt).
\]
The spectrum $\s(H)$ has multiplicity $2$ in the set
$\gS_2=\s(H)\sm\ol{\gS_4}$.
\end{theorem}

\no {\bf Remark.}
1) In the case $i_2)$ the function $\D$ is real analytic
on $(\wt\l_n^-,\wt\l_n^-+\ve)$
for some $\ve>0$. Then $\r>0$ on this interval.
Hence $\wt\l_n^-=r_m^-$ for some $m\ge 1$.
In the case $i_3)$
the similar arguments show  $\wt \l_n^+=r_m^+$ for some $m\ge 1$.
Thus, endpoints of the intervals $\s_n$ are periodic or antiperiodic
eigenvalues or resonances.

\no 2) For the small coefficients $p,q$ we have the case $i_2)$ at $n=1$
and the cases $i_1)$ for $n\ge 2$ (see Theorem \ref{1.3}
and Fig.\ref{ssp}).
The typical graph of $\D$ is given by Fig.\ref{lf}.
In Fig.\ref{lf} we have the case $i_1)$
for $n=1,2,3,5$, the case $i_3)$ for $n=4$.
For the Hill operator we have only the case $i_1)$.

\no 3) Identity \er{s4} shows
that the spectrum $\s(H)$ of $H$ has multiplicity 4 on the intervals
$\s_n'\cap\s_{n+1}'$ and $\s_n^-\cap\s_n^+$ (since two branches
$\D_1(\l),\D_2(\l)\in (-1,1)$
for all $\l\in \s_n^-\cap\s_n^+$).

\no 4) In Lemma \ref{llf} we prove that if the resonance $r$ is an
endpoint of the gap $\g$ and $\D(r)\in(-1,1)$, then $r$ has
multiplicity 1 or 2. Moreover, the spectrum $\s(H)$
near the resonance $r$ has multiplicity 4. If the gap $\g$ is
non-empty, then the resonance has multiplicity 1 and it is a branch
point of $\D$.

\no 5) For the Euler-Bernoulli equation
all resonances have multiplicities 1 or 2 and belong to $\R_-$
(see \cite{PK})
and the spectrum lies in $\R_+$ (see \cite{P1}).
The point $0$ is the unique
resonance, which is a (lowest) endpoint of the spectrum.

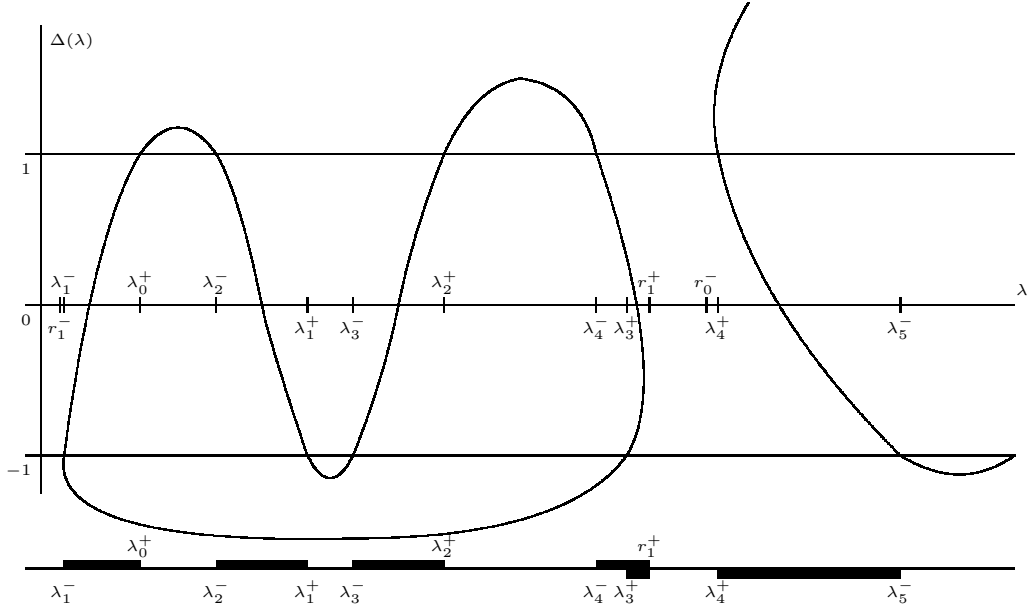
\begin{figure}
\tiny
\unitlength 1.00mm \linethickness{0.4pt}
\begin{picture}(81.00,77.00)
\put(12.00,15.00){\line(0,1){62.00}}
\put(10.00,60.00){\line(1,0){130.00}}
\put(10.00,40.00){\line(1,0){130.00}}
\put(10.00,20.00){\line(1,0){130.00}}
\put(141.00,42.00){\makebox(0,0)[cc]{$\l$}}
\put(16.00,75.00){\makebox(0,0)[cc]{$\D(\l)$}}
\put(10.00,38.00){\makebox(0,0)[cc]{$0$}}
\put(9.00,18.00){\makebox(0,0)[cc]{$-1$}}
\put(10.00,58.00){\makebox(0,0)[cc]{$1$}}
\bezier{500}(89.00,20.00)(95.00,30.00)(85.00,60.00)
\bezier{500}(85.00,60.00)(83.00,69.00)(75.00,70.00)
\bezier{500}(75.00,70.00)(69.00,69.00)(65.00,60.00)
\bezier{500}(65.00,60.00)(61.00,50.00)(59.00,40.00)
\bezier{500}(59.00,40.00)(57.00,30.00)(53.00,20.00)
\bezier{500}(53.00,20.00)(50.00,14.00)(47.00,20.00)
\bezier{500}(47.00,20.00)(42.00,35.00)(41.00,40.00)
\bezier{500}(41.00,40.00)(38.00,55.00)(35.00,60.00)
\bezier{500}(35.00,60.00)(30.00,67.00)(25.00,60.00)
\bezier{500}(25.00,60.00)(19.00,50.00)(15.00,20.00)
\bezier{500}(15.00,20.00)(13.00,8.00)(53.00,09.00)
\bezier{500}(53.00,09.00)(81.00,9.00)(89.00,20.00)
\bezier{500}(105.00,80.00)(99.00,70.00)(101.00,60.00)
\bezier{500}(101.00,60.00)(105.00,40.00)(125.00,20.00)
\bezier{500}(125.00,20.00)(133.00,15.00)(140.00,20.00)
\put(14.50,39.00){\line(0,1){2.00}}
\put(14.50,37.00){\makebox(0,0)[cc]{$r_1^-$}}
\put(15.00,39.00){\line(0,1){2.00}}
\put(15.00,43.00){\makebox(0,0)[cc]{$\l_1^-$}}
\put(25.00,39.00){\line(0,1){2.00}}
\put(25.00,43.00){\makebox(0,0)[cc]{$\l_0^+$}}
\put(47.00,39.00){\line(0,1){2.00}}
\put(47.00,37.00){\makebox(0,0)[cc]{$\l_1^+$}}
\put(35.00,39.00){\line(0,1){2.00}}
\put(35.00,43.00){\makebox(0,0)[cc]{$\l_2^-$}}
\put(92.00,39.00){\line(0,1){2.00}}
\put(92.00,43.00){\makebox(0,0)[cc]{$r_1^+$}}
\put(99.50,39.00){\line(0,1){2.00}}
\put(99.50,43.00){\makebox(0,0)[cc]{$r_0^-$}}
\put(65.00,39.00){\line(0,1){2.00}}
\put(65.00,43.00){\makebox(0,0)[cc]{$\l_2^+$}}
\put(53.00,39.00){\line(0,1){2.00}}
\put(53.00,37.00){\makebox(0,0)[cc]{$\l_3^-$}}
\put(89.00,39.00){\line(0,1){2.00}}
\put(89.00,37.00){\makebox(0,0)[cc]{$\l_3^+$}}
\put(85.00,39.00){\line(0,1){2.00}}
\put(85.00,37.00){\makebox(0,0)[cc]{$\l_4^-$}}
\put(101.00,39.00){\line(0,1){2.00}}
\put(101.00,37.00){\makebox(0,0)[cc]{$\l_4^+$}}
\put(125.00,39.00){\line(0,1){2.00}}
\put(125.00,37.00){\makebox(0,0)[cc]{$\l_5^-$}}
\put(10.00,05.00){\line(1,0){130.00}}
\put(15.00,05.50){\linethickness{3.0pt}\line(1,0){10.00}}
\put(15.00,2.00){\makebox(0,0)[cc]{$\l_1^-$}}
\put(25.00,8.00){\makebox(0,0)[cc]{$\l_0^+$}}
\put(35.00,5.50){\linethickness{3.0pt}\line(1,0){12.00}}
\put(35.00,2.00){\makebox(0,0)[cc]{$\l_2^-$}}
\put(47.00,2.00){\makebox(0,0)[cc]{$\l_1^+$}}
\put(53.00,05.50){\linethickness{3.0pt}\line(1,0){12.00}}
\put(53.00,2.00){\makebox(0,0)[cc]{$\l_3^-$}}
\put(65.00,8.00){\makebox(0,0)[cc]{$\l_2^+$}}
\put(85.00,05.50){\linethickness{3.0pt}\line(1,0){7.00}}
\put(85.00,2.00){\makebox(0,0)[cc]{$\l_4^-$}}
\put(92.00,8.00){\makebox(0,0)[cc]{$r_1^+$}}
\put(89.00,04.30){\linethickness{3.0pt}\line(1,0){3.00}}
\put(89.00,2.00){\makebox(0,0)[cc]{$\l_3^+$}}
\put(101.00,04.30){\linethickness{3.0pt}\line(1,0){24.00}}
\put(101.00,2.00){\makebox(0,0)[cc]{$\l_4^+$}}
\put(125.00,2.00){\makebox(0,0)[cc]{$\l_5^-$}}
\end{picture}
\caption{Graph of the typical function $\D$ and the corresponding
spectrum of $H$. For example, the numeric research
of the Lyapunov function
for the operator $H$ with $p=0,q=\g\sum_{n\in\Z}\d(t-n),
\g=7937.7$, shows
that the graph of this function has the similar form. } \lb{lf}
\end{figure}

We formulate our theorem about the asymptotics of the periodic and
antiperiodic eigenvalues and resonances at high energy and the
recovering the spectrum of $H$.

\begin{theorem}
 \lb{T2}
i) Each  $\s_{n}=[\l_{n-1}^+,\l_{n}^-], n\ge n_0$ for some $n_0\ge 0$,
and  the spectrum of $H$ has multiplicity $2$ in $\s_{n}$ and the intervals
$(\l_{n}^-,\l_{n}^+)\ne \es$ are gaps. Moreover, $r_n^\pm,\l_n^\pm$
satisfy:
\[
\lb{rnn}
r_n^{\pm}=-4(\pi n)^4+2p_0(\pi n)^2
\pm\sqrt2 \pi n|\wh{p_n'}|+O(1),
\]
\[
\lb{eln} \l_n^{\pm}=(\pi n)^4-\wh p_0(\pi n)^2 \pm \pi
n{|\wh{p_n'}|\/2}+O(1)
\]
as $n\to\iy$, where
$
\wh p_0=\int_0^1p(t)dt,\qq
\wh{p_n'}=\int_0^1 p'(t)e^{-i2\pi nt}dt,\qq n\ge 1.
$
If  $|\wh{p_n'}|\ge {1\/n^{\a}}$ for all large $n$ and for some $\a\in (0,1)$,
then there exists an infinite number of gaps $\g_n$ such that
$|\g_n|\to\iy$ as $n\to\iy$.

\no ii) The periodic spectrum and the antiperiodic spectrum recover
the resonances and the spectrum $\s(H)$. The periodic (antiperiodic)
spectrum and resonances recover the antiperiodic (periodic) spectrum
and $\s(H)$.
\end{theorem}

\no {\bf Remark.} \no 1) The spectrum of $H$ on $\s(H)$  has
multiplicity $2$ at high energy. The spectrum of the $2\ts 2$ matrix
Schr\"odinger operator  (see \cite{BBK}, \cite{CK}) and $4\ts 4$ Dirac operator (see \cite{K})
at high energy roughly speaking has
multiplicity $4$.

\no 2) The periodic and antiperiodic eigenvalues of our operator
accumulate   at
$+\iy$ and the resonances accumulate  at $-\iy$. In the case of the
periodic $N\ts N$ matrix Schr\"odinger operator on the real line
(see \cite{CK}) the periodic and antiperiodic eigenvalues and the
resonances accumulate at $+\iy$. In the case of $2N\ts 2N$ Dirac operator
the periodic and antiperiodic eigenvalues and the
resonances accumulate at $\pm\iy$ (see \cite{K}).

\no 3) In the case $p=0$ the asymptotics of periodic and antiperiodic
eigenvalues and resonances are determined in terms of
$\wh q_n=\int_0^1 q(t)e^{-i2\pi nt}dt$ (see \cite{BK}).

\no 4) Asymptotics \er{eln} yields that if the Fourier coefficients
$\wh{p_n'}$ of $p'$ are slowly decreasing, as $n\to\iy$,
then the spectrum $\s(H)$ has an infinite number of the gaps $\g_n$,
which increasing,  as $n\to\iy$. If $p=0$ and $q\in L^1(0,1)$,
then the gaps $\g_n$ are decreasing,  as $n\to\iy$ (see \cite{BK}).

\no 5) Under some conditions on
the matrix potential the number of gaps in the spectrum
of the matrix periodic Schr\"odinger operator (see \cite{CK}, \cite{MV})  and of the Dirac system
(see \cite{K}) is  finite.

We will show that for small potentials the lowest spectral band of
$H$ contains the interval of multiplicity 4. Below we will sometimes write
$\r(\l,p),...$ instead of $\r(\l),...$, when several
potentials are being dealt with.

\begin{theorem} \lb{1.3}
Let $H_\ve={d^4\/dt^4}+\ve {d\/dt}p{d\/dt},\ve\in\R$ and $\wh p_0=0$.
Then there exist two real analytic functions $r_0^-(\ve),
\l_0^+(\ve)$ in the disk $\{|\ve|<\ve_1\}$ for some $\ve_1>0$ such
that $r_0^{-}(\ve)$ is a simple zero of the function $\r(\cdot,\ve
p)$, and $\l_0^+(\ve)$ is a simple zero of the function
$D_+(\cdot,\ve p), r_0^-(0)=\l_0^+(0)=0$. These functions satisfy:
\[
\lb{l0} r_0^-(\ve)=2\ve^2 (4v_1-v_2)+O(\ve^3),\qq
 \l_0^+(\ve)=2\ve^2 (4v_1-v_2)+O(\ve^3),
\]
\[
\lb{bg0}
 \l_0^+(\ve)-r_0^-(\ve)=4A^2\ve^4+O(\ve^5),\ \qq
A={v_2\/12}-{4v_1\/3}=\int_0^1\rt(\int_0^tp(s)ds\rt)^2dt>0,
\]
as $\ve\to 0$, where
\[
\lb{vm} v_\n =\int_0^\n  dt\int_0^tp(s)p(t)(\n -t+s)(t-s)ds,\qq \n =1,2.
\]
Moreover, if $\ve\in (-\ve_1,\ve_1)\sm\{0\}$, then $r_0^-(\ve)<\l_0^+(\ve)$ and
 the spectrum of $H_\ve$ in the interval $(r_0^-(\ve),\l_0^+(\ve))$ has multiplicity 4.
 Other spectrum has multiplicity 2.
\end{theorem}

\no {\bf Remark.} 1) Graph of the Lyapunov function for small
coefficients is shown by Fig.\ref{ssp}.

\no 2) The similar result holds for the case $p=0$ and $q\to 0$, see
\cite{BK}.

\no 3) The spectrum  of the Euler-Bernoulli operator has
multiplicity 2 \cite{P1}. It gives that for some small specific $p\ne 0,q\ne 0$
 the spectrum of our operator $H$ has also multiplicity 2.

\begin{figure}
\tiny
\unitlength 1.00mm \linethickness{0.4pt}
\begin{picture}(81.00,77.00)
\put(12.00,8.00){\line(0,1){69.00}}
\put(10.00,61.00){\line(1,0){130.00}}
\put(10.00,41.00){\line(1,0){130.00}}
\put(10.00,21.00){\line(1,0){130.00}}
\put(141.00,43.00){\makebox(0,0)[cc]{$\l$}}
\put(7.00,75.00){\makebox(0,0)[cc]{$\D(\l)$}}
\put(10.00,38.00){\makebox(0,0)[cc]{$0$}}
\put(9.00,18.00){\makebox(0,0)[cc]{$-1$}}
\put(10.00,58.00){\makebox(0,0)[cc]{$1$}}
\bezier{100}(24.00,51.00)(24.50,61.00)(35.00,81.00)
\put(42.00,79.00){\makebox(0,0)[cc]{$\D_1(\l)$}}
\bezier{500}(24.00,51.00)(25.00,23.00)(40.00,19.00)
\bezier{500}(40.00,19.00)(45.00,17.00)(80.00,40.00)
\bezier{500}(80.00,40.00)(110.00,57.00)(130.00,62.00)
\bezier{100}(130.00,62.00)(135.00,63.00)(140.00,60.00)
\put(140.00,64.00){\makebox(0,0)[cc]{$\D_2(\l)$}}
\put(24.00,40.00){\line(0,1){2.00}}
\put(21.00,37.00){\makebox(0,0)[cc]{$r_0^-$}}
\put(26.00,41.00){\line(0,1){1.00}}
\put(29.00,44.00){\makebox(0,0)[cc]{$\l_0^+$}}
\put(35.00,40.00){\line(0,1){1.00}}
\put(38.00,37.00){\makebox(0,0)[cc]{$\l_1^-$}}
\put(48.00,40.00){\line(0,1){1.00}}
\put(46.00,37.00){\makebox(0,0)[cc]{$\l_1^+$}}
\put(127.00,40.00){\line(0,1){1.00}}
\put(129.00,37.00){\makebox(0,0)[cc]{$\l_2^-$}}
\put(138.00,40.00){\line(0,1){1.00}}
\put(137.00,37.00){\makebox(0,0)[cc]{$\l_2^+$}}
\put(24.00,42.00){\linethickness{2.0pt}\line(1,0){2.00}}
\put(24.00,40.00){\linethickness{2.0pt}\line(1,0){11.00}}
\put(48.00,40.00){\linethickness{2.0pt}\line(1,0){79.00}}
\put(138.00,40.00){\linethickness{2.0pt}\line(1,0){2.00}}
\end{picture}
\caption{The spectrum of $H^\ve$ for small $\ve$.} \lb{ssp}
\end{figure}
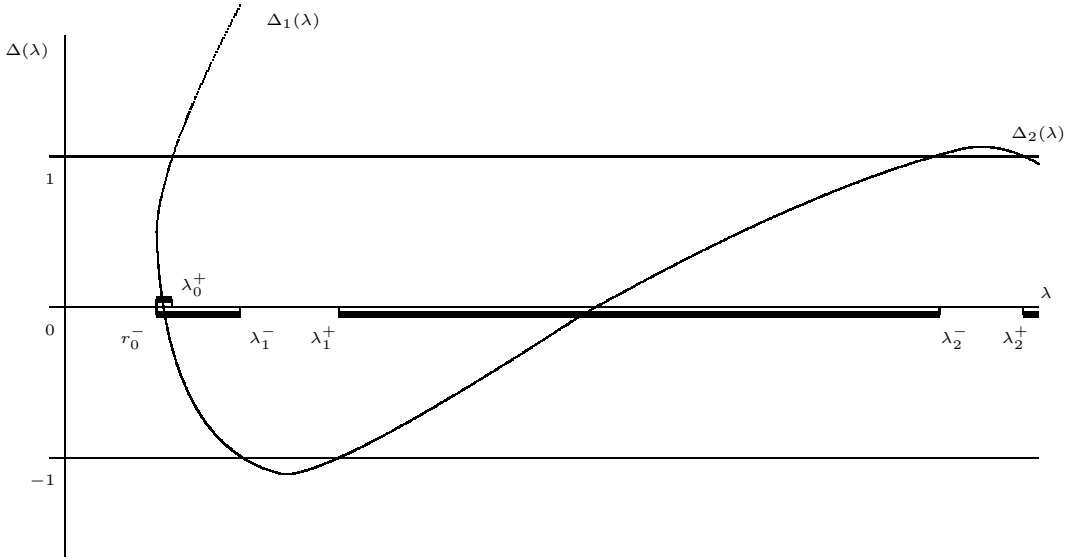

There exist many papers about the periodic systems on the real line
(see \cite{BBK}, \cite{Ca1}, \cite{Ca2}, \cite{CK}, \cite{CG},
\cite{CG1}, \cite{CHGL}, \cite{GL}, \cite{Ly}, \cite{K}, \cite{Kr},
\cite{YS}).
 The Riemann surface for the Lyapunov function
and the sharp asymptotics of the periodic and antiperiodic spectrum
and the branch points of the Lyapunov function (resonances) for the
periodic $N\ts N$ matrix Scr\"odinger operator were obtained in
\cite{CK} (in \cite{BBK} for the case $N=2$) and for the first order
systems in \cite{K}. Moreover, estimates of the  gap lengths in
terms of potentials and some new traces formulas were obtained in
\cite{CK}, \cite{K}.


The Euler-Bernoulli
equation $(ay'')''=\l b y$ with periodic coefficients $a,b$ was studied in \cite{P1}, \cite{P2}, \cite{PK}.
The spectrum lies on $\R_+$, endpoints of gaps are
periodic and antiperiodic eigenvalues, see \cite{P1}. Moreover, some inverse
problem results (similar to the Borg theorem)
were obtained in this paper.
The more detailed analysis, including the position
of the "Dirichlet" spectrum, was given in \cite{P2}.
A leading term in asymptotics
of multipliers for the $2N$ order periodic operator are detremined and an
explicit formula for the spectral expansion \cite{T3} was obtained,  the case $N=2$ see in
\cite{T1}, \cite{T2}). The asymptotic estimates for the periodic and
antiperiodic eigenvalues of the operator
$(-1)^N{d^{2N}\/dt^{2N}}+q$, where $q$ is a periodic distribution,
were obtained in \cite{MM1}, \cite{MM2}.

In \cite{BK} the following results for the operator $H$ with $p=0$
were obtained: 1)  the Lyapunov function is constructed on a
2-sheeted Riemann surface and the existence of real and complex
branch points is proved, 2) asymptotics of the periodic and antiperiodic
eigenvalues and resonances in terms of the Fourier coefficients of $q$
are determined,
3) asymptotics of $\r_0^-$ and $\l_0^+$ for small $q$ are determined.
Note that the "correct" labeling of spectral bands and gaps
is absent in \cite{BK}.

In the present paper we give the more detailed description of the spectrum,
than in \cite{BK}, for the more general operator.
We extend the results of \cite{BK} to the operator $H$ with $p\ne 0$.
Using the more careful local analysis of the Lyapunov function
we describe the spectral bands $\s_n$ in terms of this function,
including the multiplicity of the spectrum,
and determine the high energy asymptotics of bands, equipped with "correct" labeling.

Note that the operators $H={d^4\/dt^4}+{d\/dt}p{d\/dt}+q$ and
$A=8{d^3\/dt^3}+6p{d\/dt}+3p'$ constitute a Lax pair, where
$p=p(t,\t), q=q(t,\t)$ and $t$ is a space coordinate and $\t$ is a
time. The corresponding nonlinear equation $H_\t=[H,M]$ has the form
\cite{HLO}
$$
\ca
p_\t=10p'''+6pp'-24q',\\
q_\t=3(p^{(5)}+pp'''+p'p'')-8q'''-6pq'. \ac
$$

We present the plan of our paper. In Sect. 2 we obtain the basic
properties of the fundamental solutions $\vp_j, j\in \N_3^0$. These
results give the analyticity of $M$ on the complex plane. In order
to determine the asymptotics of $M(\l)$ as $|\l|\to \iy$ we define
 and study the Jost type solutions of \er{1b} (they have a good asymptotics as
$|\l|\to\iy$) in Section 3. Using the properties of the Jost type
solutions we determine the asymptotics of the monodromy matrix and
multipliers at high energy. The corresponding technical proofs are
given in Appendix. In Sect. 3 we also obtain the main properties of
the Lyapunov functions and the function $\r$. In Sect. 4 we prove
Theorems \ref{T1}-\ref{T2}. In Sect. 5 we consider the operator $H$
with the small coefficients and prove Theorem \ref{1.3}.

\section {Properties of fundamental solutions
$\vp_j,j=0,..,3$}
 \setcounter{equation}{0}

The  fundamental solutions $\vp_j^0(t,\l), \
(j,t,\l)\in\N_3^0\ts\R\ts\C$ of the unperturbed equation $y''''=\l
y$ are given by
\[
\label{21}
\vp_0^0={\cosh zt+\cos zt\/2},\ \
\vp_1^0={\sinh zt+\sin zt\/2z},\ \
\vp_2^0={\cosh zt-\cos zt\/ 2z^2},\ \
\vp_{3}^0={\sinh zt-\sin zt\/2z^3},
\]
recall that $z=x+iy=\l^{1/4},\arg z\in (-{\pi\/4},{\pi\/4}]$ and
$x\ge|y|$. Let $\|f\|=\int_0^1|f(t)|dt<\iy$.

\begin{lemma} \lb{T21}
i) Each  function $\vp_j(t,\cdot), (j,t)\in\N_3^0\ts\R_+$ is entire,
real on $\R$ and satisfies:
\[
\lb{2if} \max_{j,
k\in\N_3^0}\lt\{\lt|\l^{j-k\/4}\lt(\vp_j^{(k)}(t,\l)-
\sum_{n=0}^{N-1}\vp_{n,j}^{(k)}(t,\l)\rt)\rt|\rt\} \le {(\vk
t)^N\/N!|z|_1^N}e^{xt+\vk},\qq \vk={\|p\|}+{\|p'\|}+{\|q\|},
\]
 for all $N\ge 0,\l\in\C$, where $|z|_1=\max\{1,|z|\}$. Moreover,
 each function
 $T_\n={1\/4}\Tr M^\n, \n=1,2$, is entire, real on $\R$ and satisfies
\[
\label{2.2a}
|T_\n (\l)|\le
e^{x\n +\vk},\ \ |T_\n (\l)-T_\n ^0(\l)|\le {\n \vk\/2|z|_1}e^{x\n +\vk},\qq
\l\in\C.
\]

\no ii) Let $q=0,\wh p_0=p(0)=0$.
Then
\[
|T_\n (\l)-T_\n ^0(\l)-\e_\n (\l)|\le {(\n \vk)^3\/6|z|_1^3}e^{x\n +\vk},\qq
(\n ,\l)\in\{1,2\}\ts\C,
\label{(4.61)}
\]
where
\[
\label{(7.1)}
\e_\n (\l)=\int_0^\n
dt\int_0^tp(s)p(t)\vp_{1}^0(\n -t+s,\l)\vp_{1}^0(t-s,\l)ds.
\]
\end{lemma}

\no {\bf Proof.}
Each function $\vp_j^0(t,\cdot),(j,t)\in\N_3^0\ts\R$,
given by \er{21}, is entire and satisfies:
\[
\lb{itn}
(\vp_{j}^0)^{(k)}=\vp_{j-k}^0,\qq
\sum_{m=0}^3\vp_{j-m}^0(t)\vp_{m-k}^0(s)=\vp_{j-k}^0(t+s),\qq
|\vp_{j}^0(t)|\le {e^{tx}+e^{t|y|}\/2|z|_1^j}
\le {e^{xt}\/|z|_1^j},
\]
where $(k,s)\in\N_3^0\ts\R$. Here and below in this proof
$\vp_j^0(t)=\vp_j^0(t,\l),...$

\no i) The fundamental solutions $\vp_j$ satisfy the equation
\[
\lb{2ic} \vp_j(t,\l) =\vp_{j}^0(t)-\int_0^t\vp_{3}^0(t-s)u_j(s)ds,\
\ \ u_j=(p\vp_j')'+q\vp_j,\qq (j,t)\in\N_3^0\ts\R.
\]
Using \er{itn}, we deduce that $u_j$ satisfies the integral equation
\[
\lb{iev}
u_j(t)=u_{0,j}(t)-\int_0^tK_3(t,t-s)u_j(s)ds,\qq
u_{0,j}(t)=K_j(t,t),
\]
\[
\lb{vvj}
K_j(t,s)=p(t)\vp_{j-2}^0(s)+p'(t)\vp_{j-1}^0(s)+q(t)\vp_{j}^0(s).
\]
The standard iterations yield
\[
\lb{evj} u_j(t)=\sum_{n\ge 0} u_{n,j}(t),\ \ \ u_{n+1,j}(t)=
-\int_0^tK_3(t,s)u_{n,j}(s)ds.
\]
Substituting \er{evj} into \er{2ic} we obtain
\[
\lb{2id} \vp_j(t)=\sum_{n\ge 0}
\vp_{n,j}(t),\ \ \ \vp_{n+1,j}(t)=
-\int_0^t\vp_{3}^0(t-s)u_{n,j}(s)ds,\qq \vp_{0,j}=\vp_{j}^0
\]
and \er{evj} gives
\[
u_{n,j}(t)=(-1)^n \int\limits_{0< t_n<...< t_2< t_1\le t_0=t}
\lt(\prod\limits_{1\le k\le n}
K_3(t_{k-1},t_{k-1}-t_k)\rt)u_{0,j}(t_n)dt_1dt_2...dt_n.
\lb{2ig}
\]
Using \er{itn}, identity \er{vvj}  provides
$$
|K_j(t,s)|\le {e^{xs}\/|z|_1^{j-3}}
\lt({|p(t)|\/|z|_1}+{|p'(t)|\/|z|_1^2}+{|q(t)|\/|z|_1^3}\rt).
$$
Substituting these estimates into \er{2ig} we have
\[
\lb{fs1}
|u_{n,j}(t)|\le{(\vk t)^{n}\/|z|_1^{j+n-3}n!}e^{xt}
\lt({|p(t)|\/|z|_1}+{|p'(t)|\/|z|_1^2}+{|q(t)|\/|z|_1^3}\rt).
\]
Substituting these estimates into \er{2id}, we obtain the estimate
$$
|\vp_{n,j}(t)|
\le {(\vk t)^{n}\/|z|_1^{j+n}n!}e^{xt}\le {(\vk t)^{n}\/n!|z|_1^n}e^{xt}.
$$
This shows that for any fixed
$t\in[0,1]$ the formal series \er{2id} converges
absolutely and uniformly on
bounded subset of $\C$. Each term of this series is an entire
function. Hence the sum is an entire function. Summing the
majorants we obtain estimates  \er{2if}.

The trace of the monodromy matrix is a sum of its eigenvalues. The
set of these eigenvalues at $\l\in\R$ is symmetric with respect to
the real axis. Hence each $T_\n $ is real on $\R$.

 We will prove
\er{2.2a}. We have
\[
\lb{mum}
4T_\n =\Tr M^\n =\Tr M(\n )=\sum_{k=0}^3\vp_k^{(k)}(\n )
=\sum_{n\ge 0}f_{\n n},\qq f_{\n n}=\sum_{k=0}^3\vp_{n,k}^{(k)}(\n ),\qq \n =1,2.
\]
Identities \er{itn}, \er{2id} imply
$$
\vp_{n,k}^{(k)}(t,\l)=
-\int_0^t\vp_{3-k}^0(t-s,\l)u_{n,k}(s,\l)ds,\qq k\in\N_3^0.
$$
Estimates \er{itn}, \er{fs1} give
$|\vp_{n,k}^{(k)}(\n ,\l)|\le {(\n \vk)^n\/n!|z|_1^n}e^{x\n }$, which yields
\[
\lb{svp}
|f_{\n n}(\l)|
\le 4{(\n \vk)^n\/n!|z|_1^n}e^{x\n },\ \ n\ge 0.
\]
The last estimate shows that the series  \er{mum} converges absolutely and
uniformly on bounded subset of $\C$. Each term of this series is
an entire function. Hence the sum is an entire function and each
$T_\n $ is entire. Summing the majorants we obtain \er{2.2a}.

\no ii) Identities \er{mum}, \er{svp} yield
\[
\lb{fs3}
|T_\n -f_{\n 0}-f_{\n 1}-f_{\n 2}|\le {(\n \vk)^3\/6|z|_1^3}e^{x\n +\vk}.
\]
If we assume that
\[
\lb{fs2} f_{\n 0}=T_\n ^0,\qq f_{\n 1}=0,\qq f_{\n 2}=\e_\n ,
\]
then substituting \er{fs2} into \er{fs3} we obtain \er{(4.61)}.
 We will show \er{fs2}.

The first identity \er{fs2} follows from \er{mum}.
Identities \er{itn}, \er{iev}, \er{2id} give
$$
f_{\n 1}\!=\!-\sum_{k=0}^3\int_0^\n \!\vp_{3-k}^0(\n -t)
(p(t)\vp_{k-2}^0(t)+p'(t)\vp_{k-1}^0(t))dt
=-\vp_{1}^0(\n )\int_0^\n \!p(t)dt-\vp_{2}^0(\n )\int_0^\n \!p'(t)dt,
$$
which yields the second identity in \er{fs2}.
Substituting \er{evj} into \er{2id}
and using \er{vvj}, \er{itn} we obtain
$$
f_{\n 2}=\sum_{k=0}^3\int_0^\n dt\int_0^t
K_3(t,t-s)\vp_{3-k}^0(\n -t)(p(s)\vp_{k-2}^0(s)+p'(s)\vp_{k-1}^0(s))ds
$$
$$
=\int_0^\n dt\int_0^t(p(t)\vp_{1}^0(t-s)+p'(t)\vp_{2}^0(t-s))
(p(s)\vp_{1}^0(\n -t+s)+p'(s)\vp_{2}^0(\n -t+s))ds
$$
$$
=\!\int_0^\n \!dt\int_0^t\!(p(t)p(s)\wt\vp_{11}(t-s)+p'(t)p(s)\wt\vp_{12}(t-s)
+p(t)p'(s)\wt\vp_{21}(t-s)
+p'(t)p'(s)\wt\vp_{22}(t-s))ds,
$$
where $\wt\vp_{km}(t)=\vp_{k}^0(\n -t)\vp_{m}^0(t)$.
Integration by parts and identity $p(0)=0$ give
$$
\int_0^\n dt\int_0^tp'(t)p'(s)\wt\vp_{22}(t-s)ds
=-\int_0^\n dt\int_0^tp'(t)p(s)(\wt\vp_{12}(t-s)
-\wt\vp_{21}(t-s))ds.
$$
Then
$f_{\n 2}=\int_0^\n dt\int_0^tp(t)p(s)\wt\vp_{11}(t-s)ds+J,$
where
$$
J=\int_0^\n dt\int_0^t
(p(t)p'(s)+p'(t)p(s))\wt\vp_{21}(t-s)ds
$$
$$
=\int_0^\n dtp(t)\int_0^tp'(s)\wt\vp_{21}(t-s)ds
+\int_0^\n dsp(s)\int_s^\n p'(t)\wt\vp_{21}(t-s)dt
$$
$$
=\!\!\int_0^\n \!\!dtp(t)\int_0^tp(s)(\wt\vp_{20}(t-s)-\wt\vp_{11}(t-s))ds
-\!\int_0^\n \!\!dsp(s)\int_s^\n p(t)(
\wt\vp_{20}(t-s)-\wt\vp_{11}(t-s))dt=0.
$$
We have
$
f_{\n 2}=\int_0^\n dt\int_0^tp(t)p(s)\wt\vp_{11}(t-s)ds,
$
which yields the third identity in \er{fs2}.
$\BBox$

\section {Asymptotics}
\setcounter{equation}{0}

We introduce the diagonal matrix $\O=\O(\l)=\{\d_{kj}\o_j(\l)\}_{k,j=0}^3$,
where $\o_j=\o_j(\l)$ satisfy
\[
\lb{Om}
(\o_0,\o_1,\o_2,\o_3)
=(1,-i,i,-1),\qq\l\in\ol{\C_+},\qq
\O(\ol\l)=\ol{\O(\l)},\qq \l\in\C\sm\R.
\]
Recall that $z=\l^{1/4}, \l\in\C, \arg z\in(-{\pi\/4},{\pi\/4}]$. Identities \er{Om}
yield
\[
\Re(z\o_0)\ge\Re(z\o_1)\ge\Re(z\o_2)\ge\Re(z\o_3),\qq z=\l^{1/4},\
\arg z\in(-{\pi\/4},{\pi\/4}],\ \l\in\C.
\]
Introduce so-called Jost solutions
$\vt_j(t,\l),j\in\N_3^0,(t,\l)\in\R\ts\C$ which  satisfy equation
\er{1b} and the asymptotics $\vt_j(t,\l)=e^{z\o_jt}(1+o(1))$ as $
|\l|\to\iy$ for each fixed $t\in[0,1]$. We take $\vt_j(t,\l),
j\in\N_3^0$ which satisfies the integral equation (see \cite{N})
\[
\lb{fsv} \vt_j(t)=e^{z\o_jt}+{1\/4z^3}\int_t^1\sum_{n=0}^{j-1}\o_n
e^{z\o_n(t-s)}g_j(s)ds-{1\/4z^3}\int_0^t\sum_{n=j}^{3}\o_n
e^{z\o_n(t-s)}g_j(s)ds.
\]
where $g_j=(p\vt_j')'+q\vt_j$. Note that this equation has unique
solution  for all  $|\l|>R$ and some $R>0$ (see \cite{N}).  Let
\[
\lb{lr}
\L_r=\{\l\in\C:|\l|>r^4\max\{1,\vk^4\}\},\qq r>0,\qq
\L_r^{\pm}=\L_r\cap\C^{\pm},
\]
recall $\vk=\|p\|+\|p'\|+\|q\|$. We will prove the following lemma
in Appendix.

\begin{lemma} \lb{Lf}
The following identity holds:
\[
\lb{MY}
M=(Z\P_0)\F e^{z\O}(Z\P_0)^{-1} \qq\text{on}\qq \L_1, \qq
\text{where}\qq \F=\{\phi_{kj}\}_{k,j=0}^3=\P_0^{-1}\P_1,
\]
\[
\lb{YP}
\P_t=Z^{-1}\vT_te^{-zt\O},\qq
\vT_t=\{\vt_j^{(k)}(t,\cdot)\}_{k,j=0}^3,
\qq Z=\{\d_{kj}z^k\}_{k,j=0}^3.
\]
The functions $\phi_{kj},k,j\in\N_3^0$ are analytic in
$\L_1^{\pm}$ and satisfy:
\[
\lb{eph}
|\phi_{jj}(\l)|\le{4\/3},\qqq
|\phi_{kj}(\l)-\d_{kj}|\le{\vk\/|z|},\qqq \l\in\L_3,
\]
\[
\lb{pns}
\phi_{kj}(\l)=O(z^{-2}),\qq k\ne j,
\qq \phi_{jj}(\l)=e^{-{\o_j^3 \wh p_0\/4z}}+
O(z^{-3})\qq \text{as}\qq |\l|\to\iy,
\]
\[
\lb{ap1}
\phi_{12}(\l)=-2\x^2\ol{\wh{p_n'}}+O(\x^3),\qq
\phi_{21}(\l)=-2\x^2\wh{p_n'}+O(\x^3)\qq \text{as}
\qq z=\pi n+O(n^{-1}),
\]
\[
\lb{pbn}
\phi_{01}(\l)\phi_{10}(\l)=2|\wh{p_n'}|^2\x^4+O(\x^5)\qq \text{as}
\qq z=(1+i)\pi n+O(n^{-1}),\qq  n\to\iy,\qq \x={1\/4\pi n}.
\]
\end{lemma}

\no {\bf Remark.} 1) Identity \er{MY} shows that the matrices $\F e^{z\O}$ and $M$
has the same eigenvalues.

\no 2) If $p=q=0$, then $\F=I_4$ and $T_\n ^0={1\/4}\sum_0^3e^{z\n \o_k},\n =1,2$.

We introduce the function
\[
\lb{T}
T=4T_1^2-T_2,\qq T^0=4(T_1^0)^2-T_2^0=1+2\cosh z\cos z
={1\/2}\sum_{0\le j<k\le 3}e^{z(\o_j+\o_k)}.
\]

\begin{lemma}
\lb{res}
The functions $T_1,T$ satisfy
\[
\lb{T1T} T_1={1\/4}\sum_0^3\f_{kk}e^{z\o_k},\qq
T={1\/2}\sum_{0\le j<k\le
3}v_{jk}e^{z(\o_j+\o_k)},\qq
v_{jk}=\f_{jj}\f_{kk}-\f_{jk}\f_{kj},
\]
\[
\lb{eT}
|T_1(\l)-T_1^0(\l)|\le {\vk\/|z|}e^{x},\qq
|T(\l)-T^0(\l)|\le {9\vk\/|z|}e^{x+|y|}, \qq \l\in\L_3.
\]
\end{lemma}

\no {\bf Proof.}   Identity  \er{MY} gives $\Tr M=\Tr\F e^{z\O}$, which yields
the first identity in \er{T1T}.
Moreover, due to \er{MY}, we have
$$
T_2={1\/4}\Tr M^2
={1\/4}\Tr(\F e^{z\O})^2={1\/4}\lt(\sum_{j=0}^3\f_{jj}^2e^{2z\o_j}
+2\sum_{0\le j<k\le 3}\f_{jk}\f_{kj}e^{z(\o_j+\o_k)}\rt).
$$
Substituting this identity and the first identity
in \er{T1T} into \er{T}
we obtain the second identity in \er{T1T}.
Identities \er{T1T} give
$$
T_1-T_1^0={1\/4}\sum_{k=0}^3e^{z\o_k}(\phi_{kk}-1),\qq
T-T^0={1\/2}\sum_{0\le j<k\le 3}e^{z(\o_j+\o_k)}(v_{jk}-1).
$$
Estimates \er{eph} and $e^{z\o_k}\le e^x$
provide the first estimate in \er{eT}.
Estimates \er{eph} yield
$$
|v_{jk}-1|
\le |\f_{jj}-1|+|\f_{kk}-1|
+|\f_{jj}-1||\f_{kk}-1|+|\f_{jk}||\f_{kj}|
\le 2{\vk\/z}\lt(1+{\vk\/z}\rt)\le{9\vk\/|z|},
$$
on $\L_3$.
Using the estimates
$e^{z(\o_j+\o_k)}\le e^{x+|y|}$
we obtain the second estimate in \er{eT}.
$\BBox$

 Introduce the simply connected domains $\cD_n=\{\l\in \C:|\l^{1/4}-(1\pm i)\pi n|
<{\pi\/2\sqrt2}\},n\ge 0$, and let
$\cD=\C\sm\cup_{n\ge 0}\ol{\cD_n}$.

\begin{lemma}\lb{ar}
 i) The function $\r$, given by \er{1d}, is entire, real on $\R$ and satisfies:
\[
\lb{3a} |\r(\l)-\r^0(\l)|\le{3\vk\/|z|_1}e^{2x+\vk},\ \ \ \l\in\C,
\]
\[
\lb{ero} |\r^0(\l)|>{e^{2x}\/16},\qq \l\in\cD,\qq
\r(\l)=\r^0(\l)(1+O(\l^{-1/4})),\qq
|\l|\to\iy,\qq\l\in\cD,
\]
\[
\lb{rd}
\r={1\/16}\lt(4e^{(1-i)z}\phi_{01}\phi_{10}
+(\phi_{00}e^{z}-\phi_{11}e^{-iz})^2+O(1)\rt),\qq |\l|\to\iy,\qq
x-y\le\pi.
\]

\no ii) For each integer
$N>n_0$ for some $n_0\ge 1$ the function $\r$ has exactly $2N+1$
zeros, counted with multiplicity, in the disk
$\{\l:|\l|<4(\pi(N+{1\/2}))^4\}$ and for each $n>N$, exactly two
zeros, counted with multiplicity, in the domain
$\cD_n$. There are no other zeros.

\no iii) The function $\r$ has an odd number of real zeros,
counted with multiplicity, on the interval
$(-\G,\G)\ss\R,\G=4(\pi(N+{1\/2}))^4,N>n_0$.
\end{lemma}

\no {\bf Proof.} i) By Lemma \ref{T21}, the functions $T_\n ,\n =1,2$
are entire and real on $\R$. Then $\r$ is entire and real on $\R$. We have
\[
\lb{h1}
\r^0={T_2^0+1\/2}-(T_1^0)^2
=-\sinh^2{(1-i)z\/2}\sin^2{(1-i)z\/2}.
\]
The first identity in \er{h1} yields
$$
|\r(\l)-\r^0(\l)|
\le{|T_2(\l)-T_2^0(\l)|\/2}
+|T_1(\l)-T_1^0(\l)||T_1(\l)+T_1^0(\l)|,\qq \l\in\C.
$$
Then estimates \er{itn}, \er{2.2a} provide \er{3a}. Using \er{h1}
and the estimate $e^{|y|}<4|\sin z|$  for $|z-\pi
n|\ge{\pi\/4},n\in\Z$ (see [PT]), we obtain
$$
|\r^0(\l)|>{1\/16}e^{2|\Im{(1-i)z\/2}|+2|\Im{i(1-i)z\/2}|}
={1\/16}e^{|y+x|+|y-x|} ={e^{2x}\/16},\qq \l\in\cD,
$$
which yields the first estimate in \er{ero}. This estimate and \er{3a}
give the asymptotics in \er{ero}.

Identity \er{T1T} implies
$$
T_1(\l)={1\/4}\lt(\phi_{00}(\l)e^{z}+\phi_{11}(\l)e^{-iz}+O(e^{-x})\rt),\qq
T(\l)={1\/2}\lt(e^{(1-i)z}v_{01}(\l)+O(1)\rt)
$$
as $|\l|\to\iy,x-y\le\pi.$
The last identity in \er{ra} gives
$$
\r={1\/2}\lt(1-{1\/2}\lt[e^{(1-i)z}v_{01}+O(1)\rt]\rt)
+{1\/4^2}\lt(\phi_{00}e^{z}+\phi_{11}e^{-iz}+O(e^{-x})\rt)^2
$$
$$
={1\/16}\lt(-4e^{(1-i)z}v_{01}
+\phi_{00}^2e^{2z}+\phi_{11}^2e^{-2iz}+2\phi_{00}\phi_{11}e^{(1-i)z}+O(1)\rt)
$$
$$
={1\/16}\lt(4e^{(1-i)z}\phi_{01}\phi_{10}
+\phi_{00}^2e^{2z}+\phi_{11}^2e^{-2iz}-2\phi_{00}\phi_{11}e^{(1-i)z}+O(1)\rt)
,\qq x-y\le\pi
$$
as $ |\l|\to\iy$, which gives \er{rd}.

ii) Introduce the contour
$C_0(r)=\{\l:|\l^{1/4}|=\pi r\}$. Let $N_1>N$ be another
integer. Consider the contours
$C_0(N+{1\/2}),C_0(N_1+{1\/2}),\pa\cD_n,n>N$.
Then  \er{3a}, \er{ero} yield on all contours
$$
|\r(\l)-\r^0(\l)|\le o(1)e^{2x}<|\r^0(\l)|.
$$
Hence, by the Rouch\'e theorem, $\r(\l)$ has as many zeros,
counted with multiplicity, as $\r^0(\l)$ in each of
the bounded domains and the remaining unbounded domain. Since
$\r^0(\l)$ has exactly one simple zero at $\l=0$
and exactly one zero of multiplicity 2 at $-4(\pi n)^4,n\ge 1$,
and since $N_1>N$ can be chosen arbitrarily large,
the point ii) follows.

\no iii) The function $\r$ is real on $\R$, then
$r$ is a zero of $\r$ iff $\ol r$ is a zero of $\r$.
For large integer
$N$ the function $\r$ has exactly $2N+1$
zeros, counted with multiplicity, in the disk
$\{\l:|\l|<4(\pi(N+{1\/2}))^4\}$ and for each $n>N$, exactly two
zeros, counted with multiplicity, in the domain
$\cD_n$. There are no other zeros.
Then $\r$ has an odd number of real zeros
on the interval $(-\G,\G)$.
$\BBox$

\begin{lemma}
\lb{Dpm}
The functions $D_\pm$ are entire, real on $\R$ and satisfy:
\[
\lb{3i}
D_{\pm}={T\mp4T_1+1\/2},\qqq
|D_\pm(\l)-D_\pm^0(\l)|\le {7\vk\/|z|} e^{x+|y|},\qqq
\l\in\L_4.
\]
For each integer $N>n_0$ for some $n_0\ge 1$ the function $D_+$ has
exactly $2N+1$ zeros in the domain
$\{|\l|^{1/4}<2\pi(N+{1\/2})\}$,
the function $D_-$ has
exactly $2N$ zeros in the domain
$\{|\l|^{1/4}<2\pi N\}$,
counted with multiplicity, and for each $n>N$, the function $D_+$ has
exactly two zeros in the domain $\{|\l^{1/4}-2\pi n|<{\pi\/2}\}$,
the function $D_-$ has  exactly two zeros in the domain
$\{|\l^{1/4}-\pi(2n+1)|<{\pi\/2}\}$,
counted with multiplicity. There are no other zeros.

\end{lemma}

\no {\bf Proof.}
Identities \er{1d}, \er{2l} yield the first identity in \er{3i},
then $D_\pm$ are entire and real on $\R$.
The first identity in \er{3i} give
$$
|D_\pm-D_\pm^0|\le{|T-T^0|+4|T_1-T_1^0|\/2}
\le(9 e^{x+|y|}+ 4e^{x}){\vk\/2|z|},
$$
which yields \er{3i}.
Let $N'>N$ be another integer. Let $\l$ belong to the
contours $C_0(2N+1),C_0(2N'+1),C_{2n}({1\/2}),|n|>N$, where
$C_n(r)=\{\l:|\l^{1/4}-\pi n|=\pi r\},r>0$.  Note that
$e^{{1\/2}|y|}<4|\sin{z\/2}|,e^{{1\/2}x}<4|\sinh{z\/2}|,z=\l^{1/4},$
on all contours. Then $e^{{1\/2}(x+|y|)}<16|\sin{z\/2}\sinh{z\/2}|$
and  \er{3i} on all contours yield
$$
\lt|D_+(\l)-4\sin^2{z\/2}\sinh^2{z\/2}\rt|\le o(1)e^{x+|y|}
<\lt|4\sin^2{z\/2}\sinh^2{z\/2}\rt|.
$$
Hence, by Rouch\'e's theorem, $D_+$ has as many zeros,
as $\sin^2{z\/2}\sinh^2{z\/2}$ in each of the
bounded domains and the remaining unbounded domain. Since
$\sin^2{z\/2}\sinh^2{z\/2}$ has exactly one simple zero at $\l=0$
and exactly one zero of multiplicity two
at $(2\pi n)^4,n\ge 1$, and since $N'>N$ can be chosen arbitrarily large,
the statement for $D_+$ follows. Proof for $D_-$ is similar.
$\BBox$

\section {Proof of Theorems \ref{T1}-\ref{T2}}
\setcounter{equation}{0}

\no{\bf Proof of Theorem 1.1.}
Proof of identities \er{1d}, \er{2l} and the statements
ii), iii) repeats the arguments from \cite{BBK}, \cite{CK}.
We have only to prove \er{aD1}, \er{aD2}.
Estimates \er{eT} give
$T_1(\l)=T_1^0(\l)+e^xO(z^{-1}),|\l|\to\iy$.
Substituting this asymptotics and \er{ero} into \er{1d} we obtain
$$
\D_1(\l)=T_1^0(\l)+\sqrt{\r^0(\l)}+e^xO(z^{-1})\qq\text{as}\qq
|\l|\to\iy,\qq \l\in\cD.
$$
Using the identity $\D_1^0=T_1^0+\sqrt{\r^0}=\cosh z$ (see \er{0})
we get \er{aD1}.

Estimates \er{eT} give $T(\l)=T^0(\l)+e^{x+|y|}O(z^{-1})$ as
$|\l|\to\iy$. Substituting this asymptotics and \er{aD1} into  the
identity $\D_2={T-1\/2\D_1}$ (see \er{ra}) we obtain
$$
\D_2(\l)={T^0(\l)-1\/2\cosh z}+e^{|y|}O(z^{-1})\qq\text{as}\qq
|\l|\to\iy,\qq \l\in\cD.
$$
Using the identity $\D_2^0={T^0-1\/2\cosh z}=\cos z$, we obtain
\er{aD2}. $\BBox$

\begin{lemma}
The functions $\D_1+\D_2,\D_1\D_2$ are entire, real on $\R$ and
satisfy:
\[
\lb{ra}
\D_1^2+\D_2^2=1+T_2, \qq \D_1\D_2=2T_1^2-{T_2+1\/2}={T-1\/2},
\qq \r={1-T\/2}+T_1^0,
\]
\[
\lb{5d}
D_{\pm}=(T_!\mp 1)^2-\r={(2T_1\mp1)^2-T_2\/2},
\qq D_+-D_-=-4T_1.
\]
\end{lemma}

\no {\bf Proof.}  By Lemma \ref{T21}, the functions $T_\n ,\n =1,2$
are entire and real on $\R$. Identities \er{1d}, \er{2l} yield
\er{ra}, \er{5d}, then $\D_1+\D_2,\D_1\D_2$ are entire and real on
$\R$. $\BBox$

Below we need the following results about the Lyapunov function $\D(\l)$
in the interval $-1\le \D\le 1$ (see Fig.\ref{fl})

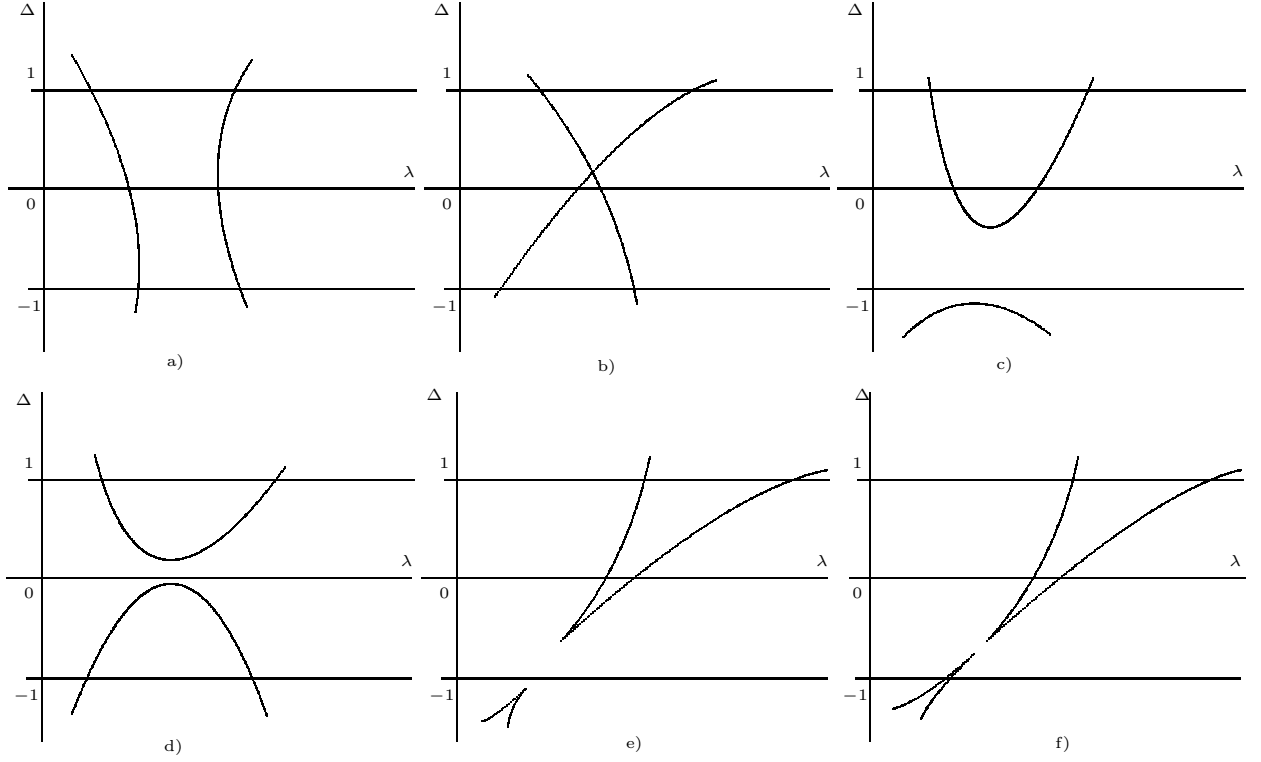
\begin{figure}
\tiny
\unitlength=1.00mm
\special{em:linewidth 0.4pt}
\linethickness{0.4pt}
\begin{picture}(165.00,95.33)
\put(5.34,2.00){\line(0,1){46.33}}
\put(0.67,23.66){\line(1,0){53.33}}
\put(3.34,10.33){\line(1,0){50.67}}
\put(3.67,36.66){\line(1,0){50.67}}
\put(53.34,26.00){\makebox(0,0)[cc]{$\l$}}
\put(3.67,21.66){\makebox(0,0)[cc]{$0$}}
\put(3.34,8.00){\makebox(0,0)[cc]{$-1$}}
\put(3.67,39.00){\makebox(0,0)[cc]{$1$}}
\put(60.00,2.00){\line(0,1){46.33}}
\put(55.33,23.66){\line(1,0){53.33}}
\put(58.00,10.33){\line(1,0){50.67}}
\put(58.33,36.66){\line(1,0){50.67}}
\put(108.00,26.00){\makebox(0,0)[cc]{$\l$}}
\put(58.33,21.66){\makebox(0,0)[cc]{$0$}}
\put(58.00,8.00){\makebox(0,0)[cc]{$-1$}}
\put(58.33,39.00){\makebox(0,0)[cc]{$1$}}
\put(5.67,53.67){\line(0,1){46.33}}
\put(1.00,75.33){\line(1,0){53.33}}
\put(3.67,62.00){\line(1,0){50.67}}
\put(4.00,88.33){\line(1,0){50.67}}
\put(53.67,77.66){\makebox(0,0)[cc]{$\l$}}
\put(4.00,73.33){\makebox(0,0)[cc]{$0$}}
\put(3.67,59.66){\makebox(0,0)[cc]{$-1$}}
\put(4.00,90.66){\makebox(0,0)[cc]{$1$}}
\put(60.33,53.67){\line(0,1){46.33}}
\put(55.66,75.33){\line(1,0){53.33}}
\put(58.33,62.00){\line(1,0){50.67}}
\put(58.66,88.33){\line(1,0){50.67}}
\put(108.33,77.66){\makebox(0,0)[cc]{$\l$}}
\put(58.66,73.33){\makebox(0,0)[cc]{$0$}}
\put(58.33,59.66){\makebox(0,0)[cc]{$-1$}}
\put(58.66,90.66){\makebox(0,0)[cc]{$1$}}
\bezier{144}(9.34,93.00)(20.34,74.00)(17.67,59.00)
\bezier{160}(32.34,59.66)(24.34,79.66)(33.00,92.33)
\bezier{140}(69.33,90.33)(80.66,76.66)(83.66,60.00)
\bezier{172}(65.00,61.00)(81.66,85.33)(94.00,89.66)
\bezier{112}(73.99,15.67)(82.33,25.33)(85.33,39.67)
\bezier{172}(73.66,15.33)(95.33,35.33)(108.66,38.00)
\bezier{236}(12.34,40.00)(19.00,13.00)(37.34,38.33)
\put(23.00,52.33){\makebox(0,0)[cc]{a)}}
\put(79.66,51.66){\makebox(0,0)[cc]{b)}}
\put(22.67,1.33){\makebox(0,0)[cc]{d)}}
\put(83.33,1.66){\makebox(0,0)[cc]{e)}}
\bezier{296}(9.34,5.66)(22.67,40.33)(35.00,5.33)
\put(3.00,47.33){\makebox(0,0)[cc]{$\D$}}
\put(3.50,99.00){\makebox(0,0)[cc]{$\D$}}
\put(58.00,99.00){\makebox(0,0)[cc]{$\D$}}
\put(57.00,48.00){\makebox(0,0)[cc]{$\D$}}
\put(114.67,53.67){\line(0,1){46.33}}
\put(110.00,75.33){\line(1,0){53.33}}
\put(112.67,62.00){\line(1,0){50.67}}
\put(113.00,88.33){\line(1,0){50.67}}
\put(162.67,77.67){\makebox(0,0)[cc]{$\l$}}
\put(113.00,73.33){\makebox(0,0)[cc]{$0$}}
\put(112.67,59.67){\makebox(0,0)[cc]{$-1$}}
\put(113.00,90.67){\makebox(0,0)[cc]{$1$}}
\put(132.00,52.00){\makebox(0,0)[cc]{c)}}
\put(112.33,99.00){\makebox(0,0)[cc]{$\D$}}
\put(114.34,2.00){\line(0,1){46.33}}
\put(111.67,23.66){\line(1,0){52.00}}
\put(112.34,10.33){\line(1,0){50.67}}
\put(112.67,36.66){\line(1,0){50.67}}
\put(162.34,26.00){\makebox(0,0)[cc]{$\l$}}
\put(112.67,21.66){\makebox(0,0)[cc]{$0$}}
\put(112.34,8.00){\makebox(0,0)[cc]{$-1$}}
\put(112.67,39.00){\makebox(0,0)[cc]{$1$}}
\bezier{112}(130.00,15.67)(138.67,25.33)(141.67,39.67)
\bezier{172}(129.67,15.33)(151.67,35.33)(163.00,38.00)
\put(139.67,1.66){\makebox(0,0)[cc]{f)}}
\bezier{52}(128.00,13.66)(121.67,7.66)(117.34,6.33)
\bezier{48}(128.00,13.66)(122.34,8.00)(121.00,5.00)
\put(113.33,48.00){\makebox(0,0)[cc]{$\D$}}
\bezier{332}(122.00,90.00)(127.33,50.33)(143.67,90.00)
\bezier{104}(118.67,55.67)(127.33,64.33)(138.00,56.00)
\bezier{28}(69.00,9.00)(64.67,5.00)(63.33,4.67)
\bezier{24}(69.00,9.00)(67.00,6.67)(66.67,4.00)
\end{picture}
\caption{Possible (a,b) and impossible (c,d,e,f) local behavior of the function $\D(\l)$
in the interval $-1\le \D\le 1$ } \lb{fl}
\end{figure}


\begin{lemma} \lb{llf}
Let $r$ be a zero of $\r$ of multiplicity $m$ and
$\D(r)\in(-1,1)$. Then  $m\le 2$.

\no i) Let $m=1$.

\no a) If $\r'(r)>0$, then
$\D(r)<\D_1(\text{or}\ \D_2)\le 1$ and
$\D_1'(\text{or}\ \D_2')>0$ on $(r,p_2)$,
and $-1\le \D_2(\text{or}\ \D_1)<\D(r)$
and $\D_2'(\text{or}\ \D_1')<0$ on $(r,a_2)$, where  $p_2$ is a periodic eigenvalue,
$a_2$ is an antiperiodic eigenvalue, and $p_2>r,a_2>r$.

\no b) If $\r'(r)<0$, then
$\D(r)<\D_1(\text{or}\ \D_2)\le 1$
and $\D_1'(\text{or}\ \D_2')<0$ on $(p_1,r)$
and $-1\le \D_2(\text{or}\ \D_1)<\D(r)$
and $\D_2'(\text{or}\ \D_1')>0$ on $(a_1,r)$,
where  $p_1$ is a periodic eigenvalue,
$a_1$ is an antiperiodic eigenvalue, and $p_1<r,a_1<r$.

\no ii) Let $m=2$. Then $\r''(r)>0$.
Moreover, $\D(r)\le\D_1(\text{or}\ \D_2)\le 1$ on $(p_1,p_2)$,
$\D_1'(\text{or}\ \D_2')>0$ on $[r,p_2)$
and $\D_1'(\text{or}\ \D_2')<0$ on $(p_1,r]$,
$-1\le\D_2(\text{or}\ \D_1)\le\D(r)$ on $(a_1,a_2)$,
$\D_2'(\text{or}\ \D_1')<0$ on $[r,a_2)$
 and $\D_2'(\text{or}\ \D_1')>0$ on $(a_1,r]$,
 where $p_1,p_2$ are periodic eigenvalues,
$a_1,a_2$ are antiperiodic eigenvalues, and $p_1<r<p_2,a_1<r<a_2$.

\no iii)
Let $\r(\l^*)>0,\D_\n(\l^*)\in(-1,1)$ for some $\l^*\in\R,\n=1,2$.
Then there exist the points $b_1,b_2$ such that
$b_1$ (or $b_2$) is a periodic and $b_2$ (or $b_1$)
is an antiperiodic eigenvalue and
exactly one from the following 3 cases holds:

$iii_1$) the function  $\D$ is real analytic and $\D'\ne 0$ on
$s=(b_1,b_2)$, $\l^*\in s$  and $\D(s)\ss(-1,1)$,

$iii_2$) there exists a zero $r_-$ of $\r$ such that
$r_-<\min \{b_1,b_2\}$ and one branch of $\D$
is real analytic and its derivative $\ne 0$ on $s_-=(r_-,b_1)$
and another branch of $\D$ is real analytic and its derivative
$\ne 0$ on $s_+=(r_-,b_2)$,

$iii_3$) there exists a zero $r_+$ of $\r$ such that
$r_+>\max \{b_1,b_2\}$ and one branch of $\D$
is real analytic and its derivative $\ne 0$ on $s_-=(b_1,r_+)$
 and another branch of $\D$ is real analytic and its
derivative $\ne 0$ on $s_+=(b_2,r_+)$.

Moreover, $\l^*\in s_-\cup s_+$ and $\D(s_-\cup s_+)\in(-1,1)$
in the cases $iii_2),iii_3)$
\end{lemma}

\no {\bf Proof.}
If $r$ is a zero of $\r$ of multiplicity $m$, then $\r(\l)=(\l-r)^m g(\l)$,
where $g$ is entire function, real on $\R$, and $g(r)\ne 0$.
Identities \er{1d} yield
$$
\D(\l)=T_1(\l)+(\l-r)^{m\/2}\sqrt{g(\l)},\qq \l\in\C,\qq |\l-r|\to 0.
$$
Consider small neighborhood
of the point $r$ and any angle in this neighborhood made by lines
originating
from the point $r$. The function $\D$ maps this angle into the angle
${m\/2}$ times bigger.
If $m\ge 3$, then the domain $\{|\D(\l)-\D(r)|<\d\}\cap\C_+$ for some
$\d>0$ has the pre-image, which is a sector with the vertex angle less
than $\pi$,
and $\D(\l)$ is real on the sides of this angle.
If, in addition, $\D(r)\in(-1,1)$, then $\D(\l)\in(-1,1)$ on the sides
of this angle.
Thus $\D(\l)\in(-1,1)$ for some non-real $\l$. By Theorem \ref{T1} iii),
these $\l$
belongs to the spectrum of $H$, which contradicts to the self-adjointness
of $H$. Hence $m\le 2$.

We will prove the statements i), ii), assuming that
if $\r(\l)>0$, then $\sqrt{\r(\l)}>0$, i.e. $\D_1(\l)\ge \D_2(\l)$.
The proof for the other case is similar.

\no i) We prove the statement a). The proof of b) is similar.
If $m=1$ and $\r'(r)>0$, then $\r<0$ on $(r-\d,r)$ and $\r>0$ on
$(r,r+\d)$ for some $\d>0$.
Identity \er{1d} shows
$\Im\D_1=\Im\D_2\ne 0$ on $(r-\d,r)$ and
$\Im\D_1=\Im\D_2=0$ on $(r,r+\d)$ for some $\d>0$.
Moreover, identity \er{1d} gives
\[
\lb{dDn}
\D_\n'=T_1'-(-1)^\n{\r'\/2\sqrt{\r}},\qq \n=1,2.
\]
The identity \er{dDn} yields
$\D_1'>0$ and $\D_2'<0$ on $(r,r+\d)$.
Due to $\D(r)\in(-1,1)$ and Theorem \ref{T1} iii), we have
$\D_1'>0$ and $\D(r)<\D_1\le 1$ on $(r,p_2)$,
where $\D_1(p_2)=1$ (then $p_2$ is a periodic eigenvalue) or $\D_1(p_2)<1$
(then $p_2$ is a resonance).

Note that $\r>0$ on $(r,p_2)$. Assume that $\D_1(p_2)<1$, i.e. $p_2$ is a resonance.
Then $p_2$ is a zero of $\r$
and $\r'<0$ on $(p_2-\d_1,p_2)$ for some $\d_1>0$.
Identity \er{dDn} gives
$\D_1'<0$ on $(p_2-\d_2,p_2)$ for some $\d_2>0$, which contradicts
$\D_1'>0$ on $(r,p_2)$.
Hence $\D_1(p_2)=1$, i.e. $p_2$ is a periodic eigenvalue.

The similar arguments show that $-1\le \D_2<\D(r)$ and
$\D_2'<0$ on $(r,a_2)$, where $a_2$ is an antiperiodic eigenvalue.

\no ii) Let $\D_1(r)=\D_2(r)\in(-1,1)$ and $r$ is a zero of $\r$ of
multiplicity
$m=2$. Then $\r(\l)=(\l-r)^2g(\l),g(r)\ne 0$, and $\r''(r)=2g(r)$.
Assume $g(r)<0$. Identity \er{1d} gives
$\D=T_1+i(\l-r)\sqrt{-g},|\l-r|\to 0.$
Consider the mapping $\D$.
The interval $(\D(r)-\ve,\D(r)+\ve)$ for some $\ve>0$ has the
pre-image
orthogonal to the real axis $\l$. Due to $\D(r)\in(-1,1)$,
we obtain $\D(\l)\in(-1,1)$ for some non-real $\l$,
which contradicts to the self-adjointness of $H$.
Thus if $\D_1(r)=\D_2(r)\in(-1,1)$ and $r$ is a zero of $\r$ of
multiplicity
$m=2$, then $g(r)>0$ and $\r''(r)>0$.
Then $\r>0$ and $\Im\D_1=\Im\D_2=0$ on $(r-\d,r+\d)$ for some $\d>0$.

Repeating the arguments from the proof of i) we obtain the other
statements.

\no iii) We consider the case $\n=1,\D_1'(\l^*)>0$.
The proof for other cases is similar.
Using Theorem \ref{T1} ii) we conclude that $\l^*\in(b,b_2)$, where
$\D_1'>0$ on $(b,b_2)$ and $b$ is a resonance  or
an antiperiodic eigenvalue and $b_2$ is a periodic eigenvalue.
If $b$ is an antiperiodic eigenvalue, then
we obtain the case $iii_1)$, where $b=b_1$.
If $b$ is a resonance, then $\D_1(b)=\D_2(b)$,
the function $\D_1'>0$ on $(b,b_2)$
and, by the statement i),ii) of this Lemma,
$\D_2'<0$ on $(b,b_2)$, where $b_2$ is an antiperiodic eigenvalue.
Then
we have the case $iii_2)$, where $b=r_-$.
$\BBox$

\no {\bf Proof of Theorem \ref{bands}.}
i) Let $\D^a=\D(\cdot,ap,aq),a\in[0,1]$. The arguments from the proof of Lemma \ref{T21}
show that the fundamental solutions $\vp_j(1,\cdot),j\in\N_3^0$, and then $T_1,\r,\D^a$,
are continuous functions of $a$ for all fixed $\l\in\C$.

The identity $\D^0=\cos\sqrt\l$ proves the statement for $a=0$.
Assume that for $a=1$ and some $n\ge 1$ the statement is incorrect.
Then there exist $a_0\in(0,1)$ such that for this $n$
and any sufficiently small $\ve>0$
the statement holds for $a=a_0-\ve$
and the statement is incorrect for $a=a_0+\ve$.
Then there exists the open interval
$\s_n'\ne\es$
with the endpoint $\l_{n-1}^+$ or $\l_n^-$
such that $\D^{a-\ve}(\s_n')\in(-1,1)$.
On the other hand, by Lemma \ref{llf} iii),
$\s_n'=\es$ for $a=a_0+\ve$.
%
%
Since $\ve>0$ is any sufficiently small number,
it contradicts to the continuity of the function $\D^a$ with
respect to $a$. Hence the statement holds for $a=1$.

\no ii) Theorem \ref{bands} i) gives that if $\l\in\s_n$ for some $n\ge 1$,
then $\D_\n(\l)\in[-1,1]$ for some $\n=1,2$.
Identity \er{sp} yields $\l\in\s(H)$. Conversely, let $\l\in\s(H)$.
Then \er{sp} shows $\D_\n(\l)\in[-1,1]$ for some $\n=1,2$.
By Lemma \ref{llf} iii), there exist
the periodic and antiperiodic eigenvalues $b_1,b_2$
such that one of the cases $iii_1)-iii_3)$ holds.
By Theorem \ref{bands} i), $b_1,b_2$ are $\l_{n-1}^+,\l_n^-$
for some $n\ge 1$ and $\l\in\s_n$.
Thus we obtain the first identity in \er{spH}.

We will prove the second identity in \er{spH}, i.e. $\s_n\cap\s_{n+2}=\es$.
We have $[\l_{n-1}^+, \l_{n}^-]\ss\s_n,
[\l_{n}^+, \l_{n+1}^-]\ss\s_{n+1},[\l_{n+1}^+, \l_{n+2}^-]\ss\s_{n+2}$.
The estimates $\l_{n-1}^+\le\l_{n+1}^-\le\l_{n+1}^+,
\l_{n}^-\le\l_{n}^+\le\l_{n+2}^-$ yield that
if $\s_n\cap\s_{n+2}\ne\es$, then $\wt\s=\s_n\cap\s_{n+1}\cap\s_{n+2}\ne\es$.
Then the function $\D$ has at least
three different values at any point $\l\in\wt\s$. Thus
we obtain the second identity in \er{spH}.

We will prove that the spectrum in $\gS_4$, given by \er{s4},
has multiplicity $4$.
If $\l$ belongs to the interior of $\s_n\cap\s_{n+1}$, then
$\D(\l)$ has two distinct values in
$[-1,1]$. By Theorem \ref{T1} iii),
the spectrum at the point $\l$ has multiplicity $4$.
Let $\l\in\s_n^-\cap\s_n^+$. Then $\D_1(\l)\ne\D_2(\l)$
and $\D_1(\l),\D_2(\l)\in[-1,1]$. By Theorem \ref{T1} iii),
the spectrum at the point $\l$ has multiplicity $4$.
Thus the spectrum in $\gS_4$ has multiplicity $4$.
If $\l\in\gS_2$, then $\D_\n(\l)\in[-1,1]$ for some unique $\n$.
Then the spectrum in $\gS_2$ has multiplicity $2$.
$\BBox$

\no {\bf Proof of Theorem \ref{T2}}.
i) By Lemma \ref{res}, the function $\r$ is real on $\R$.
Relations  \er{ero} imply $\r>0$ on $[R,+\iy)$ for some $R\in\R$.
Moreover, asymptotics \er{aD1} show
$\D_1\not\in[-1,1]$ on $[R,+\iy)$ for some $R\in\R$.
Then using asymptotics \er{aD2} and Theorem \ref{T1} ii)
we deduce that
there exists $n_0\ge 0$ such that each $\s_{n},n\ge n_0$
satisfies: $\s_{n}=(\l_{n-1}^+,\l_{n}^-)$,
the spectrum has multiplicity $2$ in $\s_{n}$ and
the intervals $(\l_{n}^-,\l_{n}^+)$ are gaps.

We will determine \er{rnn}. By Lemma \ref{ar}, $r_n^\pm\in\cD_n$ for all
sufficiently large $n\ge 1$. There exist two possibilities.
First, $\Im r_n^+=\Im r_n^-=0$. Second, $\Im r_n^+>0$, then $r_n^-=\ol{r_n^+}$.
Let $\l=r_n^+$ or $r_n^-$ in the first case, and $\l=r_n^+$ in the second case.
Then $z=\l^{1/4}=(1+i)(\pi n+\d)$,  where $\d$ satisfies
$|\d|\le 1,n>n_0$ for some $n_0\ge 1$.
Then asymptotics \er{rd} implies
\[
\lb{r1}
0=\r(\l)={e^{2\pi n+2\d}\/16}
\lt((\phi_{00}(\l)e^{i\d}-\phi_{11}(\l)e^{-i\d})^2
+4\phi_{01}(\l)\phi_{10}(\l)+O(e^{-2\pi n})\rt)
\]
as $ n\to +\iy.$
Using \er{pns} we obtain
$\phi_{jj}(\l)=1+O(n^{-1})$,
and $\phi_{kj}(\l)=O(n^{-2}),k\ne j$.
Then \er{r1} give
$\sin\d=O(n^{-1})$, which yields $\d=O(n^{-1})$.
Using asymptotics \er{pns} again we obtain
$$
\phi_{00}(\l)=e^{-{(1-i)\x\/2}\wh p_0}+O(\x^3),\qq
\phi_{11}(\l)=e^{-{(1+i)\x\/2}\wh p_0}+O(\x^3),\qq \x={1\/4\pi n}
$$
as $n\to +\iy.$
Then \er{r1} gives
$$
0=\r(\l)={e^{2\pi n+2\d-\x \wh p_0}\/4}
\lt(-\sin^2(\d+{\x \wh p_0\/2})+O(\x^4)\rt),
$$
which yields $\d=-{\x \wh p_0\/2}+\wt\d,\wt\d=O(\x^2)$.
Substituting this asymptotics and \er{pbn} into \er{r1}
we obtain
$$
0=\r(\l)={e^{2\pi n+2\d-\x \wh p_0}\/4}\lt(-\sin^2\wt\d
+2\x^4|\wh{p_n'}|^2+O(\x^6)\rt),
$$
which yields
$$
\sin\wt\d=
\pm \sqrt 2\x^2|\wh{p_n'}|+O(\x^3).
$$
Then $\wt\d=\pm \sqrt 2\x^2|\wh{p_n'}|+O(\x^3)$ and
$$
(r_n^{\pm})^{1\/4}=(1+i)(\pi n-{\x \wh p_0\/2}+\wt\d)
=(1+i)\lt(\pi n-{\x \wh p_0\/2}\mp \sqrt 2\x^2|\wh{p_n'}|
+O(\x^3)\rt),
$$
which yields \er{rnn}.

We will prove \er{eln} for $\l=\l_{2n}^\pm$. The proof for
$\l_{2n-1}^\pm$ is similar. Recall that $\l=\l_{2n}^\pm$ are periodic
eigenvalues and satisfy $\det(M(\l)-I_4)=0$. Identity \er{MY} yields
$0=\det(\F(\l)e^{z\O(\l)}-I_4)=\det(\F(\l)-e^{-z\O(\l)}),z=\l^{1\/4}$,
recall
$\O(\l)=(1,-i,i,-1),\l\in\ol{\C_+}$.
Lemma \ref{Dpm} gives $z=2\pi n+\ve$, where $\ve=\ve_{2n}^\pm$
satisfy $|\ve|<1,n>n_0$ for some $n_0\ge 1$.
Then
$$
\F-e^{-z\O}
=\ma
\f_{00}-e^{-2\pi n-\ve}&\f_{01}&\f_{02}&\f_{03}\\
\f_{10}&\f_{11}-e^{2\pi ni+i\ve}&\f_{12}&\f_{13}\\
\f_{20}&\f_{21}&\f_{22}-e^{-2\pi ni-i\ve}&\f_{23}\\
\f_{30}&\f_{31}&\f_{32}&\f_{33}-e^{2\pi n+\ve}\\
\am,
$$
here and below in this proof we write $\F=\F(\l),\f_{kj}=\f_{kj}(\l),...$ Then
$$
\det(\F-e^{-z\O})=e^{2\pi n}\det\ma
\f_{00}-e^{-2\pi n-\ve}&\f_{01}&\f_{02}&\f_{03}\\
\f_{10}&\f_{11}-e^{2\pi ni+i\ve}&\f_{12}&\f_{13}\\
\f_{20}&\f_{21}&\f_{22}-e^{-2\pi ni-i\ve}&\f_{23}\\
e^{-2\pi n}\f_{30}&e^{-2\pi n}\f_{31}&e^{-2\pi n}\f_{32}&
e^{-2\pi n}\f_{33}-e^{\ve}\\
\am.
$$
Using the first estimate in \er{eph} we obtain
$$
\det(\F-e^{-z\O})=-e^{2\pi n+\ve}\lt(\det\ma
\f_{00}-e^{-2\pi n-\ve}&\f_{01}&\f_{02}\\
\f_{10}&\f_{11}-e^{2\pi ni+i\ve}&\f_{12}\\
\f_{20}&\f_{21}&\f_{22}-e^{-2\pi ni-i\ve}\\
\am+O(e^{-2\pi n})\rt)
$$
which yields
\[
\lb{pei} 0=\det (\F-e^{-z\O})=-\phi_{00}e^{2\pi n+\ve}(F_0+F_1),\qq
F_0=\det\ma
\phi_{11}-e^{i\ve}&\phi_{12}\\
\phi_{21}&\phi_{22}-e^{-i\ve}
\am,
\]
$$
F_1=-{\phi_{01}\-\phi_{00}}\det\ma
\phi_{10}&\phi_{12}\\
\phi_{20}&\phi_{22}-e^{-i\ve}
\am
+{\phi_{02}\/\phi_{00}}\det\ma
\phi_{10}&\phi_{10}-e^{i\ve}\\
\phi_{20}&\phi_{21}
\am+O(e^{-2\pi n}).
$$
Asymptotics \er{pns} gives
\[
\lb{zaF1} F_0=\det\ma
e^{-{i\wh p_0\/4z}}-e^{i\ve}+O(n^{-3})&O(n^{-2})\\
O(n^{-2})&e^{{i\wh p_0\/4z}}-e^{-i\ve}+O(n^{-3})
\am,\qq F_1=O(n^{-4}),
\]
$\l=\l_{2n}^\pm$, which yields
$F_0=2-2\cos(\ve+{\wh p_0\/4z})+O(n^{-3})$. Identity
$F_0+F_1=0$ gives
$\ve=-{\x \wh p_0\/2}+\wt\ve,\wt\ve=O(\x^{3\/2}),\x={1\/4\pi n}$.
Moreover, $e^{-{i\wh p_0\/4z}}-e^{i\ve}=O(\x^{{3\/2}})$ and
$e^{{i\wh p_0\/4z}}-e^{-i\ve}=O(\x^{{3\/2}})$.
We obtain
$F_0=2-2\cos(\ve+{\wh p_0\/4z})+O(\x^4)$, then
$\wt\ve=O(n^{-2})$.
Asymptotics \er{pns} provides
$\phi_{11}(\l)-e^{i\ve}=O(\x^2),
\phi_{22}(\l)-e^{-i\ve}=O(\x^2)$.
Substituting \er{pns}, \er{ap1} into \er{pei} we obtain
$F_1=O(n^{-6})$ and
$$
F_0=\det\ma
e^{-{i\wh p_0\/4z}}-e^{i\ve}+O(\x^3)
&-{1\/2}\x^2\ol{\wh{p_{2n}'}}+O(\x^3)\\
-{1\/2}\x^2\wh{p_{2n}'}+O(\x^3)
&e^{{i\wh p_0\/4z}}-e^{-i\ve}+O(\x^3)
\am
$$
$$
=\det\ma
e^{-{i\x \wh p_0\/2}}(1-e^{i\wt\ve})
+O(\x^3)
&-{1\/2}\x^2\ol{\wh{p_{2n}'}}+O(\x^3)\\
-{1\/2}\x^2\wh{p_{2n}'}+O(\x^3)
&e^{{i\x \wh p_0\/2}}(1-e^{-i\wt\ve})
+O(\x^3)
\am
$$
$$
=2-2\cos\wt\ve+
\lt(\sin{\x \wh p_0\/2}+\sin(\wt\ve-{\x \wh p_0\/2})\rt)O(\x^3)
-{\x^4\/4} |\wh{p_{2n}'}|^2+O(\x^5)
=\wt\ve^2+\wt\ve O(\x^3)
-{\x^4\/4}|\wh{p_{2n}'}|^2+O(\x^5).
$$
The identity $F_0+F_1=0$ gives
$
\wt\ve=\pm{\x^2\/2} |\wh{p_{2n}'}|
+O(\x^3).
$
Then
$$
(\l_{2n}^{\pm})^{1\/4}=2\pi n-{\x \wh p_0\/2}
\pm {\x^2\/2}|\wh{p_{2n}'}|
+O(\x^3),
$$
which implies \er{eln} for $\l_{2n}^{\pm}$.

\no ii) Using the identities \er{ra}, \er{5d}, asymptotics
\er{rnn}, \er{eln} and repeating the standard arguments from \cite{BK},
based on the Hadamard factorizations of the entire functions $D_\pm,\r$,
we obtain the needed statements.
$\BBox$

\section {The spectrum for the small potential}
\setcounter{equation}{0}

\no {\bf Proof of Theorem \ref{1.3}.}  The arguments from the proof of
Lemma \ref{T21} show that each function $T_\n (\l,\ve p),\n =1,2$
is entire  in $(\l,\ve)\in\C^2$ for fixed $p$. Then
$T_\n ^\ve(\l)=T_\n (\l,\ve p)$ and  $\r^\ve(\l)=\r(\l,\ve p),
D_\pm^\ve(\l)=D_\pm(\l,\ve p)$ are entire functions of $\ve$. Using
$\wh p_0=0, p'\in L_{loc}^2(\R)$ we can assume $p(0)=0$. Lemma \ref{T21}
gives
\[
\lb{Tg} T_\n ^\ve(\l)=T_\n ^0(\l)+\ve^2\e_\n (\l)+O(\ve^3),\qq \ve\to 0,
\]
uniformly on any bounded subset of $\C$.
Substituting \er{Tg} into \er{1d} we obtain
\[
\lb{re} \r^\ve(\l)=\r^0(\l)+\ve^2\wt\r(\l,\ve),
\qq\wt\r(\l,\ve)={\e_2\/2}-2T_1^0(\l)\e_1(\l)+O(\ve),\qq \ve\to 0,
\]
uniformly in any bounded domain in $\C$. The function $\r^0$ has
simple zero $\l=0$ and $\wt\r$ is analytic at the point
$(\l,\ve)=(0,0)$. Applying the Implicit Function Theorem to
$\r^\ve=\r^0+\ve^2 \wt\r$ and ${\pa\/\pa\l}\r^\ve|_{\l=\ve=0}\ne 0$,
we obtain a unique solution $r_{0}^-(\ve),|\ve|<\ve_1,r_{0}^-(0)=0$
of the equation $\r^\ve(\l)=0,|\ve|<\ve_1$ for some $\ve_1>0$.

Substituting \er{Tg} into  \er{5d} we obtain
\[
\lb{wD}
 D_+^\ve(\l)
=D_+^0(\l)+\ve^2 \wt D_+(\l,\ve),\qq \wt
D_+(\l,\ve)=2(2T_1^0(\l)-1)\e_1(\l)-{\e_2(\l)\/2}+O(\ve),
\]
as $\ve\to 0$, uniformly in any bounded domain in $\C$. The function
$D_+^0$ has simple zero $\l=0$ and $\wt D_+$ is analytic at the
point $(\l,\ve)=(0,0)$. Applying the Implicit Function Theorem to
$D_+^\ve=D_+^0+\ve^2 \wt D_+$ and
${\pa\/\pa\l}D_+^\ve|_{\l=\ve=0}\ne 0$, we obtain a unique solution
$\l_{0}^+(\ve),|\ve|<\ve_1,\l_{0}^+(0)=0$ of the equation
$D_+^\ve(\l)=0,|\ve|<\ve_1$ for some $\ve_1>0$.

We determine asymptotics \er{l0}, \er{bg0}.
Identities \er{0}, \er{(7.1)} yield
\[
\lb{ze} T_\n ^0(\l)
=1+{\n ^4\/4!}\l+O(\l^2),\qq
\e_\n (\l)=v_\n +O(\l),\qq |\l|\to 0,\qq \n =1,2,
\]
where $v_\n $ are given by \er{vm}. Let $\l=r_0^-(\ve)$. Identity
\er{0} gives $\r^0(\l)={\l\/4}+O(\l^2),\ve\to 0$. Substituting  this
asymptotics into the first identity in \er{re} we obtain
$0=\r^\ve(\l)={\l\/4}+O(\l^2)+O(\ve^2),\ve\to 0$, which yields
$\l=O(\ve^2)$. Then $\r^0(\l)={\l\/4}+O(\ve^4)$, and substituting
\er{ze} into the second asymptotics in \er{re} we obtain
$\wt\r(\l,\ve)={v_2\/2}-2v_1+O(\ve), \ve\to 0$. Substituting these
asymptotics into the first identity in \er{re} again we obtain
$$
0=\r^\ve(\l)={\l\/4}+\ve^2\lt({v_2\/2}-2v_1\rt)+O(\ve^3),\qqq\l=r_0^-(\ve),
\qqq \ve\to 0,
$$
which yields the first asymptotics in \er{l0}.

Let $\l=\l_0^+(\ve)$. Identities \er{0} imply
$D_+^0(\l)=-{\l\/4}+O(\l^2),\ve\to 0$ and the first identity \er{wD}
gives $0=D_+^\ve(\l)=-{\l\/4}+O(\l^2)+O(\ve^2)$, which yields
$\l=O(\ve^2)$. Then $D_+^0(\l)=-{\l\/4}+O(\ve^4)$ and substituting
\er{ze} into the second asymptotics in \er{wD} we obtain $\wt
D_+(\l,\ve)=2v_1-{v_2\/2}+O(\ve)$. Substituting these asymptotics
into the first identity in  \er{wD} we have
$$
0=D_+^\ve(\l)=-{\l\/4}+\ve^2\lt(2v_1-{v_2\/2}\rt)+O(\ve^3),\qqq\l=\l_0^+(\ve),\qqq\ve\to
0,
$$
which yields the second asymptotics in \er{l0}.

We prove \er{bg0}.
Asymptotics \er{l0}, \er{re} give
\[
\lb{rg0} \r^\ve(\l_0^+)=sy(\ve),\ \ \
y(\ve)=(\r^\ve)'(r_0^-)+O(s)=(\r^0)'(r_0^-)+O(\ve^2)
={1\/4}+O(\ve^2)\ \ \text{as} \ \ \ve\to 0,
\]
where $s=\l_0^+-r_0^-\to 0$. Substituting  $\r^\ve(\l_0^+)=sy(\ve)$
into the identity $D_+=(T_1-1)^2-\r$ (see \er{5d}), and using
$D_+(\l_0^+)=0$ we obtain
\[
\lb{ers} s=\l_0^+-r_0^-={(T_1^\ve(\l_0^+)-1)^2\/y(\ve)}.
\]
Substituting asymptotics \er{l0} into \er{Tg} and using \er{ze},
we obtain
\[
\lb{T1g} T_1^\ve(\l_0^+)=1-\ve^2A+O(\ve^3),\qq
A={v_2\/12}-{4v_1\/3}, \qq  \ve\to 0.
\]
Substituting \er{rg0}, \er{T1g}  into \er{ers}
we have \er{bg0}.

Recall the identity  $\D_\n ^\ve=T_1^\ve-(-1)^\n \sqrt{\r^\ve},\n =1,2$.
Then
$$
\D_\n ^\ve(\l)=T_1^\ve(r_0^-)-(-1)^\n
\sqrt{\l-r_0^-}\sqrt{y(\ve)}+O((\l-r_0^-)^{3\/2}), \qq \l-r_0^-\to
+0.
$$
Hence the function $\D_1^\ve$ is increasing and $\D_2^\ve$ is
decreasing in some interval $(r_0^-,r_0^-+\ve)$, $\ve>0$ (see
Fig.(\ref{ssp})). Asymptotics \er{bg0}, \er{T1g} give
$$
\D_1^\ve(r_0^-)=\D_2^\ve(r_0^-)=T_1^\ve(r_0^-)
=T_1^\ve(\l_0^+)+O(\ve^4)=1-\ve^2A+O(\ve^3)\qq\text{as}\qq
 \ve\to 0.
$$
Below we will prove that $A>0$. Then there exists $\d>0$ such that
$-1<\D_1^\ve(r_0^-)<1$ for each $\ve:-\d<\ve<\d$. Due to $\D_1^\ve$
is increasing in $(r_0^-,r_0^-+\ve)$, $\ve>0$ and Theorem 1.1 iv),
$\D_1^\ve$ is increasing in the interval $(r_0^-,\l_0(\ve))$, where
$\D_1^\ve(\l_0(\ve))=1$. Hence $\l_0(\ve)=\l_{2n}^{\pm}(\ve)$ for
some $n$. Note that $\l_0(0)=0$, since $\D_1^0(\l)=\cosh z$. Recall
$\l_0^+(0)=0$. Then $\l_0(\ve)=\l_0^+(\ve)$. Hence $-1<\D_1^\ve<1$
on $\a=(r_0^-(\ve),\l_0^+(\ve))$ and $\D_1^\ve(\l_0^+)=1$. Moreover,
substituting asymptotics \er{Tg}, \er{re} into the identities
$\D_2^\ve=T_1^\ve-\sqrt{\r^\ve}$, we obtain $\D_2^\ve=\cos
z+o(\ve),\ve\to 0$. Then the function $\D_2^\ve+1,-\d<\ve<\d$ has no
any zero in the interval $\a$. Then $-1<\D_2^\ve<1$ on $\a$. By
Theorem 1.1 iii), the spectrum in the interval $\a$ has multiplicity
4.

Now we will show that $A>0$.
Using \er{vm} direct calculations give
\[
\lb{A} A=\int_0^1 f(u)\int_u^1 p(t)p(t-u)dt du,\qq
f(u)=u(u-1).
\]
We have
\[
f(t)=\sum_n f_ne^{i2\pi nt},\ \ \ f_n={2\/(2\pi n)^2}, \ \ n\ne 0,\ \ \
f_0=-{1\/6},\ \  p(t)=\sum_n p_ne^{i2\pi nt}.
\]
Substituting these identities into \er{A} we get
$$
A=\int_0^1 f(s)\int_s^1 p(t)p(t-s)dt ds
=\sum_{m,n} p_np_m\int_0^1 f(s)ds\int_s^1
e^{i2\pi (n+m)t}e^{-i2\pi ns}dt=F_1+F_2,
$$
where
$$
F_2=\sum_{m+n\ne 0} {p_np_m\/2\pi i(n+m)}
\int_0^1 f(s)e^{-i2\pi ns}(1-e^{i2\pi (n+m)s})ds
=\sum_{m+n\ne 0} p_np_m{f_n-f_m\/2\pi i(n+m)}=0
$$
and
$$
F_1=\sum_{-\iy}^\iy |p_n|^2\int_0^1 f(s)(1-s)e^{-i2\pi ns}ds.
$$
Note that
$$
\int_0^1 f(s)(1-s)e^{-i2\pi ns}ds
=\sum_k \int_0^1(1-s) f_ke^{i2\pi (k-n)s}ds
=\sum_{k\ne n} {-f_k\/i2\pi (k-n)}+{f_n\/2}.
$$
Then
$$
F_1=\sum_{n\ne 0}|p_n|^2\lt(\sum_{k\ne n}{-f_k\/i2\pi (k-n)}+{f_n\/2}\rt)
=\sum_{n\ne 0}|p_n|^2{f_n\/2}>0,
$$
since $\wh p_0=0, F_1$ is real, $f_k>0,k\ne 0.\BBox$

\section {Appendix}
\setcounter{equation}{0}

We will solve \er{fsv} in terms of $f_j=z^{-2}e^{-zt\o_j}g(\cdot,\vt_j)$.
Each $f_j,j\in\N_3^0$ satisfies the equation
\[
\lb{ief}
f_{j}=
v_j+{1\/4z}\sum_{n=0}^3\o_n v_n w_{nj},
\qqq w_{nj}(t,\l)=\int_0^1e_{nj}(t-s,\l)f_j(s,\l)ds,
\]
where
$
v_j=\o_j^2p+{\o_jp'\/z}+{q\/z^2},
$
\[
\lb{e}
e_{kj}(t,\l)=\ca e^{zt(\o_k-\o_j)}\chi(-t),\ \text{if}\ k<j\\
-e^{zt(\o_k-\o_j)}\chi(t),\ \text{if}\ k\ge j
\ac,\qqq \chi(t)=\ca 0,\ \text{if}\ t<0\\ 1,\ \text{if}\ t\ge 0\ac.
\]
If $(k,j,t)\in(\N_3^0)^2\ts\R$ and $(j-k)t\ge 0$, then
$|e^{z t\o_j}|=e^{t\Re(z\o_j)}\le e^{t\Re(z\o_k)}.$
Hence
\[
\lb{ee}
|e_{kj}(t,\l)|\le 1,\qqq (k,j,t,\l)\in(\N_3^0)^2\ts\R\ts\C.
\]

\begin{lemma}  \lb{T31}
For each $(j,\l)\in\N_3^0\ts\L_1$
the integral equation \er{ief}
has the unique solution $f_j(\cdot,\l)\in L^1(0,1)$.
Each function $f_{j}(t,\cdot),t\in [0,1]$
is analytic in $\L_1^\pm$ and satisfies
\[
\lb{efj}
\|f_j(\cdot,\l)\|\le 2\vk,\qqq (j,\l)\in\N_3^0\ts\L_2.
\]

\end{lemma}

\no {\bf Proof.}
Iterations in \er{ief} provide the identities
\[
\lb{ef}
f_j=\sum_{n=0}^\iy f_{j,n},\qq f_{j,0}=v_j,
\qq f_{j,n}(t,\l)={1\/4z}\int_0^1K_j(t,s,\l)f_{j,n-1}(s,\l)ds,\ \ n\ge 1,
\]
where $K_j(t,s,\l)=\sum_0^3\o_nv_n(t,\l)e_{nj}(t-s,\l).$
Identity \er{ef} give
$$
f_{j,n}(t,\l)={1\/(4z)^n}\int_{[0,1]^n}K_j(t,t_n,\l)K_j(t_n,t_{n-1},\l)...
K_j(t_2,t_1,\l)f_{j,0}(t_1,\l)dt_1...dt_n.
$$
Using \er{ee} we have
$$
\max \{{1\/4}|K_j(t,s,\l)|,|f_{j,0}(t,\l)|\}
\le |p(t)|+{|p'(t)|\/|z|}+{|q(t)|\/|z|^{2}},\qq
(t,s,\l)\in [0,1]^2\ts\C,
$$
and
$$
\|f_{j,n}(\cdot,\l)\|
\le{1\/|4z|^n}\int_{[0,1]^{n+1}}|K_j(t,t_n,\l)||K_j(t_n,t_{n-1},\l)|...
|K_j(t_2,t_1,\l)||f_{j,0}(t_1,\l)|dt_1...dt_ndt
$$
\[
\lb{esf}
\le{1\/|z|^n}\lt(\|p\|+{\|p'\|\/|z|}+{\|q\|\/|z|^{2}}\rt)^{n+1}
\le{\vk^{n+1}\/|z|^n},\ \ \ |z|>1.
\]
These estimates show that for each
fixed $\l\in\L_1$ series
\er{ef} converges absolutely and uniformly on the interval $[0,1]$. Hence it
gives the unique solution of equation \er{ief}.
For each $t\in\R$
the series \er{ef} converges absolutely and uniformly on any bounded subset of
$\L_1$. Each term of this
series is an analytic function of $\l$ in $\L_1^\pm$. Hence
for each fixed $t\in[0,1]$ the
function $f_j(t,\cdot)$ is analytic in $\l\in\L_1^\pm$.
Summing the majorants we obtain
$
\|f_j\|\le {\vk\/1-{\vk\/|z|}},
$
which yields \er{efj}.
$\BBox$

We will show that
\[
\lb{phi}
\phi_{kj}(\l)=\d_{kj}+{\o_k\/4z}w_{kj}(\chi_{k-j},\l)s_{k-j}
-{\o_k\/(4z)^2}\sum_{k<r\le
3}\o_rw_{kr}(0,\l)w_{rj}(\chi_{r-j},\l)s_{r-j}
$$
$$
+{\o_k\/(4z)^3}\sum_{k<n<r\le 3} \o_r\o_nw_{kn}(0,\l)w_{nr}(0,\l)
w_{rj}(\chi_{r-j},\l)s_{r-j} -{\o_3w_{3j}(1,\l)\/(4z)^4}\prod_{0\le
r\le 2}\o_rw_{r,r+1}(0,\l),
\]
$\l\in\L_1$, where
$\chi_n=\ca 0,\ \text{if}\ n<0\\
1\ \text{if}\ n\ge 0\ac\!\!\!\!\!,\
s_n=\ca -1\ \text{if}\ n< 0\\
1\ \text{if}\ n\ge 0\ac$.

\no {\bf Proof of Lemma \ref{Lf}.}
The matrix $\cM(t,\l)=\{\vp_j^{(k)}(t,\l)\}\}_{k,j=0}^3$ satisfies
\[
\lb{Me} \cM'=\left(\begin{array}{cccc}
0&1&0&0\\
0&0&1&0\\
0&0&0&1\\
\l-q &-p'&-p&0\\
\end{array}\right)\cM,\qq (t,\l)\in[0,1]\ts\C,\qq \cM(0,\l)=I_4.
\]
The matrix $\vT$ is a solution of \er{Me} and
$\cM(t,\cdot)=\vT_t\vT_0^{-1}, t\in[0,1].$
Using \er{YP} we obtain
$
M=\cM(1,\cdot)=Z\P_1e^{z\O}(Z\P_0)^{-1},
$
which yields \er{MY}.

By Lemma \ref{T31}, $\P$ is analytic in $\L_1^{\pm}$,
then $\Phi $ is also analytic.
Identity \er{YP} for $\P_t=\{\p_{kj}(t,\cdot)\}_{k,j=0}^3$ implies
$\p_{kj}=z^{-k}\vt_j^{(k)}e^{-zt\o_j},$ and
identities \er{fsv}, \er{ief} show
$
\p_{kj}=\o_j^k+{1\/4z}\sum\limits_{n=0}^3\o_n^{k+1}w_{nj},
$
which yields
$$
\P_t=X+{X\O W_t\/4z},\qq X=\{\o_j^k\}_{k,j=0}^3,
\qq W_t=\{w_{kj}(t,\cdot)\}_{k,j=0}^3.
$$
The last identity in \er{MY} implies
$$
\Phi=(X^{-1}\P_0)^{-1}X^{-1}\P_1
=\lt(I_4+{\O W_0\/4z}\rt)^{-1}\lt(I_4+{\O W_1\/4z}\rt).
$$
Identities \er{e} give
\[
\lb{w=0}
w_{kj}(0,\l)=0,\qq k\ge j,\qqq
w_{kj}(1,\l)=0,\qq k<j.
\]
Hence $(\O W_0)^4=0$ and
$
(I_4+{\O W_0\/4z})^{-1}=\sum_0^3(-1)^n({\O W_0\/4z})^n.
$
Then
$$
\Phi=I_4+\lt({\O\/4z}-{\O W_0\O\/(4z)^2}
+{(\O W_0)^2\O\/(4z)^3}\rt)(W_1-W_0)-{(\O W_0)^3\O W_1\/(4z)^4}.
$$
Identities \er{w=0} yield
$w_{kj}(1,\l)-w_{kj}(0,\l)=w_{kj}(\chi_{k-j},\l)s_{k-j},$
which implies \er{phi}.

Estimates \er{ee}, \er{efj} give
$\|w_{nj}(\cdot,\l)\|\le 2\vk,\l\in\L_2.$
Then \er{phi} implies
$$
|\phi_{kj}(\l)-\d_{kj}|\le{\vk\/2z}+(3-k)\lt({\vk\/2z}\rt)^2
+{(4-k)(3-k)\/2}\lt({\vk\/2z}\rt)^3+\lt({\vk\/2z}\rt)^4.
$$
Since $|z|\ge 3\vk$ for $\l\in\L_3^{\pm}$, we have the second
estimate in \er{eph}. Using this estimate, we obtain
$|\phi_{jj}(\l)|\le 1+|\phi_{jj}(\l)-1|\le 1+{\vk\/|z|}$, which yields
the first estimate in \er{eph}.

Let us denote
$(E_{kj}f)(t,\l)=\int_0^1e_{kj}(t-s,\l)f(s)ds$ for the function
$f\in L^1(0,1)$, where $e_{kj}$ are given by \er{e}.
Let $k\ne j$. If $f'\in L^1(0,1)$, then integration by parts gives
\[
\lb{e1}
(E_{kj}f)(t,\l)
={1\/z(\o_k-\o_j)}\lt(f(t)-e^{z(\o_k-\o_j)(t-\chi_{j-k})}f(\chi_{j-k})
+(E_{kj}f')(t,\l)\rt).
\]
Identity \er{ief} yields
\[
\lb{ief1}
f_{j}=v_j+{\o_j\/4z}v_jw_{jj}+{1\/4z}\sum_{n\ne j}\o_nv_nw_{nj},
\qq v_j=\o_j^2p+{\o_jp'\/z}+{q\/z^2},\qq w_{kj}=E_{kj}f_j.
\]
Substituting \er{ief1} into the last identity and using \er{e1}
we obtain $w_{kj}(\chi_{k-j},\l)=O(z^{-1})$ as $|\l|\to\iy$.

Identity \er{phi} imlpies
\[
\lb{fkj}
\f_{kj}(\l)=\d_{kj}+{\o_k\/4z}w_{kj}(\chi_{k-j},\l)s_{k-j}+O(z^{-4})
\qq\text{as}\qq |\l|\to\iy,\qq k,j\in\N_3^0.
\]
This yields the first asymptotics in \er{pns}.
Let $z=(1+i)\pi n+O(n^{-1})$.
Then $z(\o_0-\o_1)=z(1+i)=-2i\pi n+O(n^{-1})$,
$(E_{01}(v_kw_{k1}))(0,\l)=O(n^{-1}),k\in\N_3^0$,
and \er{e1}, \er{ief1} give
$$
w_{01}(0,\l)=(E_{01}f_1)(0,\l)
=\o_1^2(E_{01}p)(0,\l)+{\o_1\/z}(E_{01}p')(0,\l)+O(n^{-2})
$$
$$
={\o_0\o_1(E_{01}p')(0,\l)\/z(\o_0-\o_1)}
+O(n^{-2})=-2\x\wh{p_n'}+O(\x^2).
$$
Substituting this identity into \er{fkj} we get
$\phi_{01}(\l)=(1-i)\x^2\wh{p_n'}+O(\x^3)$ as
$z=(1+i)\pi n+O(n^{-1}), n\to\iy.$
The similar arguments imply
$\phi_{10}(\l)=(1+i)\x^2\ol{\wh{p_n'}}+O(\x^3)$,
which yields \er{pbn}. The proof of \er{ap1} is similar.

We will prove the second asymptotics in \er{pns}.
Identities \er{ief}, \er{e}, \er{ief1} provide
$$
w_{jj}(1,\l)=-\int_0^1 f_j(t,\l)dt=-\o_j^2\wh p_0+{\o_j\wh p_0^2\/8z}
+O(z^{-2})\qq\text{as}\qq |\l|\to\iy.
$$
Substituting this asymptotics into \er{fkj} we obtain
$$
\phi_{jj}(\l)=1+{\o_j\/4z}w_{jj}(1,\l)+O(z^{-4})
=1-{\o_j^3\wh p_0\/4z}+{\o_j^2\wh p_0^2\/32z^2}+O(z^{-3}),
$$
which yields the second asymptotics in \er{pns}.
$\BBox$

\end{document}